\documentclass[aps,prd,preprint,groupedaddress,longbibliography,nofootinbib]{revtex4-2}
\usepackage[utf8]{inputenc}
\usepackage{amsmath,amssymb}
\usepackage{amsfonts}
\usepackage{slashed}
\usepackage{xcolor}

\usepackage[pdftex]{hyperref}
\hypersetup{ 
pdfstartview=FitV,
colorlinks=true, 
linkcolor=black, 
citecolor=blue, 
filecolor=black, 
urlcolor=magenta,
}
\usepackage{graphicx}
\usepackage{changes}

\begin{document}

\preprint{CPHT-089.112020,  IFT-UAM/CSIC-20-165}

\title{Hydrodynamic diffusion and its breakdown near AdS$_2$ quantum critical points}

\author{Daniel Are\'an}
\email{daniel.arean@uam.es}
\affiliation{Instituto de F\'isica Te\'orica UAM/CSIC and
Departamento de F\'isica Te\'orica, Universidad Aut{\'o}noma de Madrid\\
Campus de Cantoblanco, 28049 Madrid, Spain}
\author{Richard A. Davison}
\email{r.davison@hw.ac.uk}
\affiliation{Department of Mathematics and Maxwell Institute for Mathematical Sciences, Heriot-Watt University, Edinburgh EH14 4AS, U.K.}
\author{Blaise Gout\'eraux}
\email{blaise.gouteraux@polytechnique.edu}
\author{Kenta Suzuki}
\email{kenta.suzuki@polytechnique.edu}
\affiliation{CPHT, CNRS, Ecole polytechnique, IP Paris, F-91128 Palaiseau, France}

\date{\today}

\begin{abstract}
Hydrodynamics provides a universal description of interacting quantum field theories at sufficiently long times and wavelengths, but breaks down at scales dependent on microscopic details of the theory. {In the vicinity of a quantum critical point, it is expected that some aspects of the dynamics are universal and dictated by properties of the critical point}. We use gauge-gravity duality to investigate the breakdown of diffusive hydrodynamics in two low temperature states dual to black holes with AdS$_2$ horizons{, which exhibit quantum critical dynamics with an emergent scaling symmetry in time}. We find that the breakdown is characterized by a collision between the diffusive pole of the retarded Green's function with a pole associated to the AdS$_2$ region of the geometry, such that the local equilibration time is set by infra-red properties of the theory. The absolute values of the frequency and wavevector at the collision ($\omega_{eq}$ and $k_{eq}$) provide a natural characterization of all the low temperature diffusivities $D$ of the states via $D=\omega_{eq}/k_{eq}^2$ where $\omega_{eq}=2\pi\Delta T$ is set by the temperature $T$ and the scaling dimension $\Delta$ of an operator of the infra-red quantum critical theory. We confirm that these relations are also satisfied in an SYK chain model in the limit of strong interactions. {Our work paves the way towards a deeper understanding of transport in quantum critical phases.}
\end{abstract}

%\keywords{}

\maketitle
\newpage
\tableofcontents

\vskip1cm

\section{Introduction}

Interacting quantum field theories are notoriously challenging, especially when there is no quasiparticle-based description of the state. To describe the late time, long wavelength dynamics of these states, one can instead rely on effective approaches such as hydrodynamics. This approach has been used to gain insight into the quark-gluon plasma \cite{Kovtun:2004de,Heller:2016gbp,Romatschke:2016hle,Berges:2020fwq}, ultracold atomic systems \cite{Enss2019} and electronic transport in metals \cite{Hartnoll:2007ih,Hartnoll:2014lpa,Levitov_2016,Bandurin_2016,Crossno_2016,Moll_2016,Lucas:2017idv}. 

At long times and wavelengths, hydrodynamics provides an effective description of a system in terms of a few conserved quantities dictated by symmetries \cite{forster1975hydrodynamic,Kovtun:2012rj,Hartnoll:2016apf}. Their evolution is governed by local conservation equations for the densities and associated currents, along with constitutive relations expressing the currents in terms of the densities in a gradient expansion. The late time relaxation of the system back to equilibrium is governed by the hydrodynamic modes: poles of the retarded Green's functions of the densities with gapless dispersion relations \cite{kadanoff_hydrodynamic_2000}.

While extremely powerful, hydrodynamics breaks down at sufficiently short scales set by the local equilibration time and length. At such scales, the dynamics of the system can no longer be truncated to just the evolution of the conserved densities. Additional degrees of freedom play a significant role, and appear as additional poles of the retarded Green's function with lifetimes comparable to those of the hydrodynamic modes.\footnote{\baselineskip1pt They may also manifest themselves as branch cuts, as is the case in some weakly coupled quantum field theories \cite{Hartnoll:2005ju,Romatschke:2015gic,Kurkela:2017xis}.} In cases where the density response exhibits a parametrically slow mode arising due to a weakly broken symmetry, the breakdown of hydrodynamics manifests itself as a collision in the complex frequency plane at nonzero wavevector $k_{eq}$ between the hydrodynamic mode and the slow mode. In this case, it is often possible to augment the hydrodynamic description to incorporate this slow mode \cite{Hartnoll:2007ih,Davison:2014lua,Lucas:2015vna,Lucas:2015lna,Davison:2016hno,Delacretaz:2017zxd,Chen:2017dsy,Davison:2018ofp,Grozdanov:2018fic,Davison:2018nxm}. But typically the modes relevant for the breakdown of hydrodynamics are not of this nature, and a more complete knowledge of a system's microscopic details is required to understand them. 

In contrast to this, in the non-zero temperature quantum critical phases found near quantum phase transitions, transport properties are typically universal and are governed by scaling properties of the critical point \cite{sachdevbook}. We might then expect that the breakdown of hydrodynamics in a quantum critical phase displays a greater degree of universality. However, it is often challenging to calculate concrete observables beyond thermodynamics due to the lack of analytically tractable models without any simplifying limits, such as a large number of degrees of freedom.

In this work, we exploit gauge-gravity duality to address hydrodynamic transport and its breakdown in quantum critical phases. {Gauge-gravity duality maps} the late time dynamics of certain large $N_c$ quantum field theories (where $N_c$ is the rank of the gauge group) to theories of gravity with a negative cosmological constant \cite{Ammon:2015wua,Zaanen:2015oix,Hartnoll:2016apf}. The relaxation of conserved densities back to equilibrium is captured exactly by the evolution of perturbations of asymptotically anti de Sitter (AdS) black holes, which can be studied to obtain a precise understanding of the breakdown of hydrodynamics and the modes responsible for it \cite{Horowitz:1999jd,Son:2002sd}.

Even in the absence of a weakly broken symmetry, the breakdown of hydrodynamics can be characterized by an energy scale $\omega_{eq}$ and wavenumber $k_{eq}$, which are sensitive to the system's microscopic details. $\omega_{eq}$ and $k_{eq}$ are defined as the absolute values of the complex frequency $\omega$ and complex wavenumber $k$ at which the hydrodynamic pole of the retarded Green's function first collides with a non-hydrodynamic pole or branch point \cite{Withers:2018srf,Grozdanov:2019kge,Grozdanov:2019uhi,Heller:2020hnq}.\footnote{\baselineskip1pt {A different mechanism for the breakdown of hydrodynamics arises when interactions between hydrodynamic modes are included, in the guise of a non-analytic frequency dependence of the retarded Green's functions} (see e.g.~\cite{Kovtun:2012rj} for a review). {This mechanism is} expected to be suppressed in the large $N$ limit \cite{Kovtun:2003vj}.} The convergence properties of the real-space hydrodynamic gradient expansion {in the linear regime} are governed by $k_{eq}$  \cite{Heller:2020uuy},\footnote{\baselineskip1pt {See \cite{Heller:2013fn} for a study of  the convergence of nonlinear hydrodynamics in real time.}} which also coincides with the radius of convergence of the small-$k$ expansion of the hydrodynamic dispersion relation $\omega_{\text{hydro}}(k)$. See \cite{Abbasi:2020ykq,Jansen:2020hfd,Grozdanov:2020koi,Choi:2020tdj} for recent applications of this.

In this work, we study the breakdown of hydrodynamics in certain low temperature ($T$) states dual to black holes with nearly-extremal AdS$_2\times$R$^2$ near-horizon metrics. {These are examples of quantum critical phases with an emergent scaling symmetry in time \cite{Liu:2009dm,Cubrovic:2009ye,Faulkner:2009wj,Iqbal:2011in}, which has been dubbed `semi-local quantum critical'.} {Formally, this scaling symmetry corresponds to an infinite Lifshitz scaling exponent, $z=+\infty$: time scales but space does not.} Such states are closely related to the Sachdev-Ye-Kitaev (SYK)-like models of electrons in strange metals, which are governed by the same type of infra-red fixed point in the limit of large number of fermions and strong interactions \cite{SY92,Sachdev:2010um,Almheiri:2014cka,Kitaevtalk,Sachdev:2015efa,Maldacena:2016hyu,Maldacena:2016upp,Jensen:2016pah,Engelsoy:2016xyb,Gu:2016oyy,Davison:2016ngz,Sachdev:2019bjn}.\footnote{\baselineskip1pt The dimensionless SYK coupling is the interaction strength over temperature, so strong interactions are equivalent to low temperatures.} Specifically, we study the AdS$_4$ neutral, translation-breaking black brane of \cite{Bardoux:2012aw,Andrade:2013gsa} and the AdS$_4$-Reissner-Nordström (AdS$_4$-RN) black brane, and the breakdown of the hydrodynamics governing the diffusive transport of energy, charge and momentum in their dual states. The states we are interested in do not include any slow modes in the sense described above. Instead, local equilibration is controlled by the {intrinsic dynamics of the quantum critical degrees of freedom of AdS$_2\times$R$^2$}. We identify simple, general results for the local equilibration scales $\omega_{eq}$ and $k_{eq}$ and confirm that these also apply to the SYK chain model studied in \cite{Choi:2020tdj} in the limit of strong interactions.

Our first result is that the breakdown is caused by modes associated to the AdS$_2$ region of the geometry, and as a consequence $\omega_{eq}$ is set by universal (i.e.~infra-red) data via
\begin{equation}
\label{eq:relaxationtimeintro}
\omega_{eq}\rightarrow 2\pi \Delta T\;\;\;\text{as}\;\;\;T\rightarrow0,
\end{equation}
where $\Delta$ is the infra-red scaling dimension of the least irrelevant operator that couples to the diffusion mode. This is in contrast to systems with a weakly broken symmetry, for which $\omega_{eq}\ll T$,{but is in line with the expectation that the quantum critical dynamics is controlled by a `Planckian' timescale $\tau_{eq}\sim1/T$ \cite{sachdevbook,zaanen2004superconductivity}}. More precisely, we find that at small $k$ and $T$ the Fourier space locations of the longest-lived non-hydrodynamic poles are inherited from infra-red Green's functions, and are {approximately} located at {$\omega_n\equiv-i(n+\Delta)2\pi T$} for non-negative integers $n$. The breakdown is characterized by a collision, parametrically close to the imaginary $\omega$ axis, between the $n=0$ mode (which has a weak $k$-dependence) and the hydrodynamic mode. This collision manifests itself as a branch-point singularity in the dispersion relation of the mode.

Secondly, we find that at low temperatures the corrections to the quadratic approximation $-iDk^2$ to the exact hydrodynamic dispersion relation are parametrically small such  that the collision occurs when $k$ is almost real and
\begin{equation}
\label{eq:diffusivityintro}
k_{eq}^2\rightarrow \frac{\omega_{eq}}{D}\;\;\;\text{as}\;\;\;T\rightarrow0.
\end{equation}
In other words, the scales $k_{eq}$ and $\omega_{eq}$ governing the regime of validity of hydrodynamics are set simply by the diffusivity $D$ and the scaling dimension $\Delta$. In some of the examples we study (those involving diffusion of energy), the relevant diffusivity is controlled by an irrelevant deformation of the AdS$_2$ fixed point and in these cases the result \eqref{eq:diffusivityintro} indicates that $k_{eq}$ is controlled by the same irrelevant deformation. A priori, the result \eqref{eq:diffusivityintro} is quite surprising: it relates the radius of convergence to just the leading order term in the hydrodynamic expansion. This is a consequence of the AdS$_2$ fixed point.

By rearranging equation \eqref{eq:diffusivityintro} we obtain an answer to the question raised in \cite{Hartnoll:2014lpa} of what the underlying velocity and time scales are that govern the diffusivity in non-quasiparticle systems. In all our examples they are set by the local equilibration scales
\begin{equation}
\label{eq:Dscalesintro}
D\rightarrow v_{eq}^2\tau_{eq},\;\;\;\text{as}\;\;\; T\rightarrow0,
\end{equation}
where $v_{eq}\equiv\omega_{eq}/k_{eq}$ and $\tau_{eq}\equiv\omega_{eq}^{-1}$ are the velocity and timescale associated to local equilibration.

In the cases where diffusive hydrodynamics breaks down due to a parametrically slow mode protected by a weakly broken symmetry, $D$ is typically set by $\tau_{eq}$ and the speed of the propagating mode that dominates following the breakdown.\footnote{\label{footnoteslowmode}\baselineskip1pt Unlike here, $\tau_{eq}$ is often defined by the lifetime of a $k=0$ mode. In our conventions, these examples have $D\rightarrow v_{eq}^2\tau_{eq}/2$ in the limit of slow relaxation.} We emphasize that the breakdown of hydrodynamics is qualitatively different in the cases we study: there is not a single slow mode but a tower of AdS$_2$ modes with parametrically similar lifetimes set by $\omega_n$, and the breakdown does not produce a propagating mode with velocity $v\simeq v_{eq}$. More generally, the local equilibration time has been argued to set an upper bound on the diffusivity in \cite{Hartman:2017hhp,Lucas:2017ibu}. All examples that we study are consistent with a bound of the form $D\lesssim v_{eq}^2\tau_{eq}$ for the range of parameters we have investigated.

In the absence of a slow mode, it was proposed that low temperature diffusivities are set by the butterfly velocity $v_B$ and Lyapunov time $\tau_L$ that characterize the onset of scrambling following thermalization of the system \cite{Blake:2016wvh}. This was shown to robustly apply to the diffusivity of energy density $D_\varepsilon$ in holographic theories and SYK-like models \cite{Blake:2016sud,Gu:2016oyy,Davison:2016ngz,Blake:2016jnn,Blake:2017qgd,Guo:2019csw}. For the examples we study, $\tau_L^{-1}=2\pi T$ and
\begin{equation}
\label{eq:Dchaosintro}
D_{\varepsilon}\rightarrow v_B^2\tau_L\;\;\;\text{as}\;\;\; T\rightarrow0,
\end{equation}
which is furthermore true in general for states governed by an infra-red AdS$_2$ with the universal deformation \cite{Blake:2016jnn}.

As for our result \eqref{eq:Dscalesintro}, equation \eqref{eq:Dchaosintro} can be viewed as a consequence of the excellent applicability of the quadratic approximation to the exact hydrodynamic dispersion relation up to the relevant scale. Specifically, pole-skipping analysis suggests that the energy diffusion mode satisfies $\omega_{\text{hydro}}(k=iv_B^{-1}\tau_L^{-1})=i\tau_L^{-1}$ \cite{Grozdanov:2017ajz,Blake:2017ris,Blake:2018leo}, from which \eqref{eq:Dchaosintro} follows assuming corrections to the quadratic, diffusive form $-iD_{\varepsilon}k^2$ at $k=iv_B^{-1}\tau_L^{-1}$  are parametrically small as $T\rightarrow0$.

Our result \eqref{eq:Dscalesintro} is more general than \eqref{eq:Dchaosintro} in that it is true for all diffusivities in the examples we study, not just the diffusion of energy. Fundamentally this is because, by definition, all diffusive modes pass through the location set by $(\omega_{eq},k_{eq})$, while only the energy diffusion mode satisfies the pole-skipping constraint above \cite{Blake:2019otz}. As a consequence we provide a new perspective on, and generalization of, the relations between equilibration, transport and scrambling and their applications in AdS$_2$/SYK-like models of electrons in strange metals.

In the remainder of this work, we explain how we arrive at equations \eqref{eq:relaxationtimeintro} and \eqref{eq:diffusivityintro} before closing with comments on implications and the more general applicability of our results.

\section{Diffusive hydrodynamics}

The spectrum of hydrodynamic modes is dependent on the system under consideration, and by our definition each hydrodynamic mode has its own associated local equilibration scales $\omega_{eq}$ and $k_{eq}$. We will focus on hydrodynamic diffusion modes, which arise when a system has a current density $j$ with constitutive relation
\begin{equation}
j(\rho)=-D_\rho\nabla\rho+O(\nabla^3),
\end{equation}
where $\rho$ is the corresponding conserved density. $\rho$ will then obey the diffusion equation with diffusivity $D_\rho$ at leading order in the derivative expansion, and has a retarded Green's function \cite{kadanoff_hydrodynamic_2000,Kovtun:2012rj}
\begin{equation}
G_{\rho\rho}(\omega,k)=\frac{D_\rho\chi_{\rho\rho} k^2+\ldots}{-i\omega+D_\rho k^2+\ldots},
\end{equation}
where $\chi_{\rho\rho}\equiv\lim_{\omega\to0}G_{\rho\rho}(\omega,k)$ is the static susceptibility of $\rho$, and ellipses denote terms with higher powers of $\omega$ and $k$. The dispersion relation of the hydrodynamic diffusion mode is then
\begin{equation}
\label{eq:hydrodiffusion}
\omega_{\text{hydro}}(k)=-iD_\rho k^2+O(k^4).
\end{equation}
Hydrodynamic diffusion is a very general phenomenon. Even within the restricted class of systems that we study, the set of conserved densities that exhibit hydrodynamic diffusion varies. In this work, we will be interested in the diffusion of energy $\varepsilon$ and of transverse momentum $\Pi$.

\section{Diffusion in a neutral holographic state}

We begin with the AdS$_4$ neutral translation-breaking model \cite{Bardoux:2012aw,Andrade:2013gsa} which is a classical solution of the action
\begin{equation}
S=\int d^4x\sqrt{-g}\left(R+6-\frac{1}{2}\sum_{i=1}^{2}\left(\partial\varphi_i\right)^2\right),
\end{equation}
with spacetime metric
\begin{equation}
\label{eq:planarmetric}
ds^2=-r^2f(r)dt^2+r^2\underline{dx}^2+\frac{dr^2}{r^2f(r)},
\end{equation}
supported by two scalar fields $\varphi_i=mx^i$ $(i=1,2)$ that break translational symmetry. The emblackening factor of the solution is
\begin{equation}
\label{eq:axionemblackeningfunc}
f(r)=1-\frac{m^2}{2r^2}-\left(1-\frac{m^2}{2r_0^2}\right)\frac{r_0^3}{r^3},
\end{equation}
where $r_0$ denotes the location of the horizon with associated temperature $T$. The linear perturbations can be written in terms of four decoupled variables and we will focus on the one exhibiting the single hydrodynamic mode of the system. See Appendix \ref{sec:AxionAppendix} for further details on this spacetime, and on the calculations leading to the results below.

\subsection{Hydrodynamic mode}

The hydrodynamic mode corresponds to diffusion of energy, with the small $k$ dispersion relation \eqref{eq:hydrodiffusion} and diffusivity $D_\varepsilon\rightarrow\sqrt{3/2}m^{-1}$ in the low $T$ limit \cite{Davison:2014lua}.

First we quantify corrections to this result that will enable us to understand the breakdown of hydrodynamics. In the low temperature limit $T\sim k^2\sim\epsilon\ll 1$, the retarded Green's function of energy density $G_{\varepsilon\varepsilon}$ exhibits a pole located at
\begin{equation}
\begin{aligned}
\label{eq:scalingdispersion}
\omega(k)=-i\epsilon\sqrt{\frac{3}{2}}\frac{k^2}{m}\left(1+\epsilon\frac{k^2}{m^2}+\epsilon^2\left(\frac{4\pi T^2}{3m^2}+\frac{k^4}{m^4}\right)+\ldots\right),
\end{aligned}
\end{equation}
where we have explicitly written all $\epsilon$ dependence. For suitably small $k$, this is an approximation to the {dispersion relation of the hydrodynamic mode} {as we show in Figure \ref{fig:axion2modesmaintext}}. It becomes invalid near specific wavenumbers $k^2=k_n^2$ related to the breakdown of hydrodynamics, which will be addressed shortly.

It is important to note that \eqref{eq:scalingdispersion} is different than the hydrodynamic expansion: corrections to the quadratic $k^2$ term are not being neglected as in the usual gradient expansion, but are parametrically small in this limit under consideration. One consequence of this is that if we define any wavenumber $k_*$ with $k_*^2\sim T$ at low $T$, and define $\omega_*$ to be the location of the hydrodynamic pole at this wavenumber, then \eqref{eq:scalingdispersion} implies $D_\varepsilon\rightarrow i\omega_*/k_*^2$ as $T\rightarrow0$. For example, choosing $k_*=iv_B^{-1}\tau_L^{-1}$ results in the chaos relation \eqref{eq:Dchaosintro} as described in the Introduction.\footnote{\baselineskip1pt See Figure 2 of \cite{Blake:2018leo} for a visual representation of this.}

We will soon show that the breakdown of hydrodynamics at low $T$ is characterized by a pole collision at $k_{eq}^2\sim T,\omega_{eq}\sim T$ and thus the diffusivity can alternatively be expressed simply in terms of these scales by \eqref{eq:diffusivityintro}. But prior to exploring the pole collision that characterizes the breakdown of hydrodynamics, it is instructive to first understand the origin of the non-hydrodynamic mode responsible.

\subsection{Infra-red modes}

At low $T$ the state is governed by an infra-red fixed point manifest in the emergence of a near-horizon AdS$_2\times$R$^2$ metric with SL($2$,R) symmetry (see Appendix \ref{sec:AxionAppendix}). Each linear perturbation of the spacetime can be characterized by $\Delta(k)$, a wavenumber-dependent scaling dimension of the corresponding operator with respect to this infra-red fixed point, and a corresponding infra-red Green's function \cite{Faulkner:2009wj,Hartnoll:2012rj}
\begin{equation}
\label{eq:IRgreensfn}
\mathcal{G}_{IR}\propto T^{2\Delta(k)-1}\frac{\Gamma\left(\frac{1}{2}-\Delta(k)\right)\Gamma\left(\Delta(k)-\frac{i\omega}{2\pi T}\right)}{\Gamma\left(\frac{1}{2}+\Delta(k)\right)\Gamma\left(1-\Delta(k)-\frac{i\omega}{2\pi T}\right)}.
\end{equation}
For the spacetime perturbation that exhibits a diffusive mode, $\Delta(k)=(1+\sqrt{9+8k^2/m^2})/2$.

\begin{figure}
\begin{center}
\includegraphics[width=0.7\textwidth]{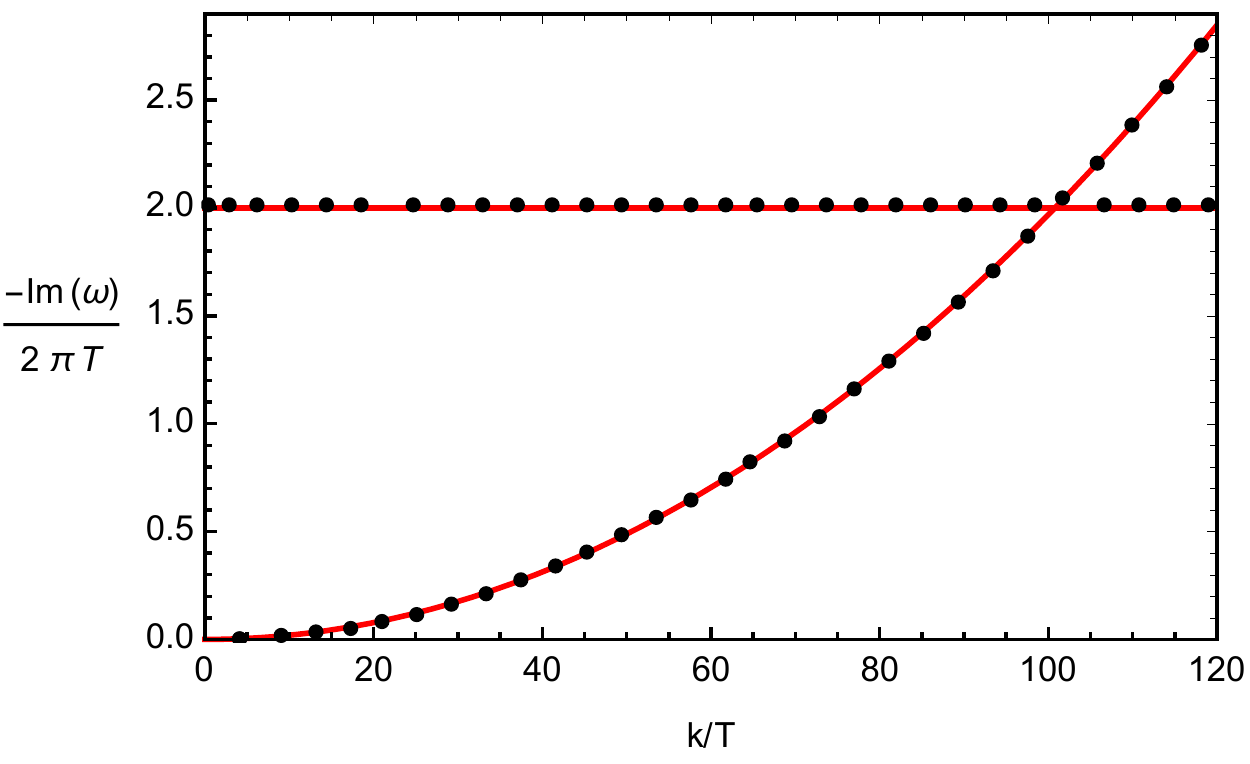}
\end{center}
\caption{Frequencies of the hydrodynamic and longest lived infra-red modes at $T/m=10^{-3}$. Black circles are numerical results and red lines are the analytic expressions \eqref{eq:scalingdispersion} and \eqref{eq:IRpolelocations}. {For real $k$, all poles displayed have purely imaginary frequencies.} }
\label{fig:axion2modesmaintext}
\end{figure}

Although analytically reconstructing $G_{\varepsilon\varepsilon}$ from $\mathcal{G}_{IR}$ is not easy, for our purposes it is enough to observe that $G_{\varepsilon\varepsilon}$ exhibits poles whose locations approach those of the poles of $\mathcal{G}_{IR}$ as $k,T\rightarrow0$. Specifically, this means that in this limit $G_{\varepsilon\varepsilon}$ exhibits poles at
\begin{equation}
\label{eq:IRpolelocations}
{\omega\rightarrow}\omega_n=-i2\pi T(n+\Delta(0)),\quad\quad\quad n=0,1,2,\ldots,
\end{equation}
with $\Delta(0)=2$. {The dispersion relation of the $n=0$ pole is shown in Figure \ref{fig:axion2modesmaintext}}.

At low $T$, the infra-red modes \eqref{eq:IRpolelocations} have a parametrically longer lifetime than the other non-hydrodynamic poles of $G_{\varepsilon\varepsilon}$, and are responsible for the breakdown of hydrodynamics.

The wavenumbers at which our calculation of the low $T$ dispersion relation \eqref{eq:scalingdispersion} of the hydrodynamic mode is invalid are $k_n^2=\sqrt{8/3}(2+n)\pi mT+O(T^2)$,\footnote{{See \eqref{eq:kndefn} for a precise expression for $k_n$.}} for which the mode would be located at precisely $\omega(k_n)=\omega_n$ in the limit of low $T$. The natural interpretation would therefore be that the invalidity of the calculation at these values of $k^2$ can be traced to the nearby presence of the infra-red mode (assuming that the location of the infra-red mode has a weak $k$-dependence). A more refined calculation below confirms this, as well as the existence of a collision between these modes for complex $k$ that signals the breakdown of hydrodynamics.

\subsection{Breakdown of hydrodynamics}

\begin{figure}
\begin{center}
\includegraphics[width=0.7\textwidth]{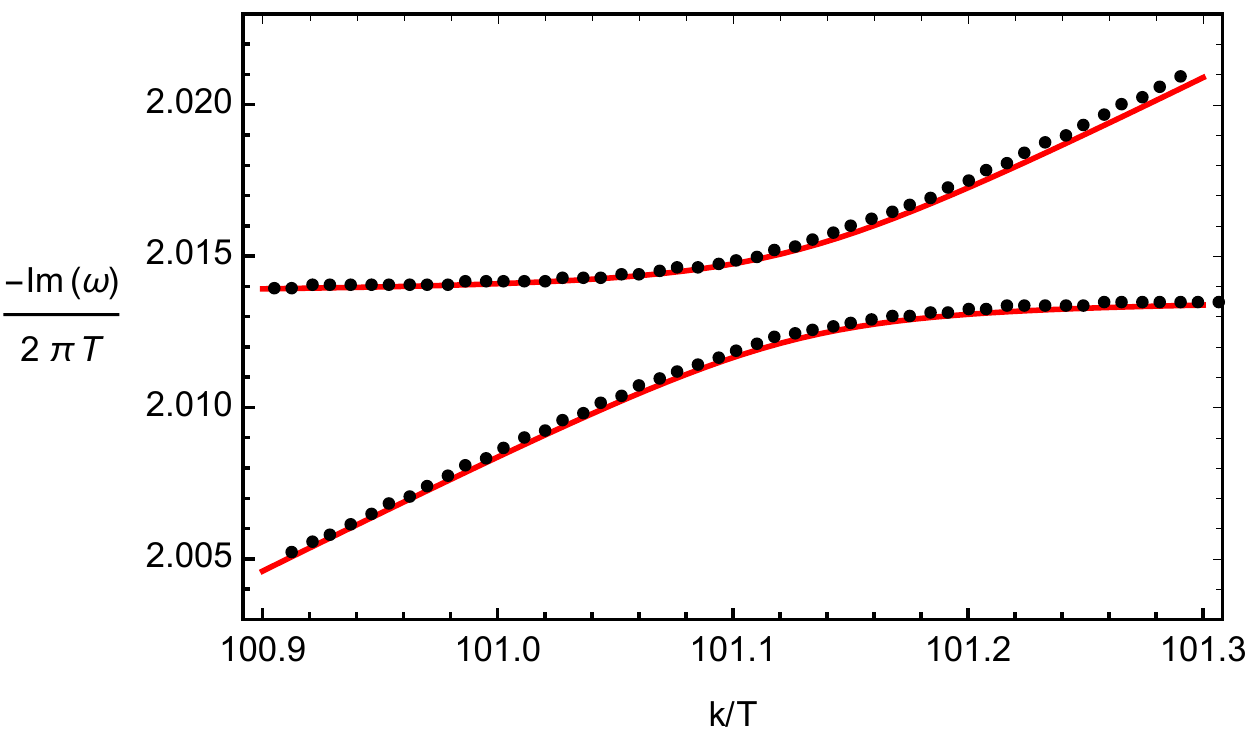}
\end{center}
\caption{{Frequencies of the hydrodynamic and longest lived infra-red modes at $T/m=10^{-3}$, zooming in on the region near $\omega_0$. The black dots are the numerical results (the bottom dots are the hydrodynamic mode, the top ones the longest-lived non-hydrodynamic mode), the red lines show the dispersion relations extracted analytically from \eqref{eq:Gnearcollision} for real values of $k$. The pole collision is not visible on this figure as it happens at a complex value of $k$. }}
\label{fig:axioncollision}
\end{figure}

\begin{figure}
\begin{center}
\includegraphics[width=0.7\textwidth]{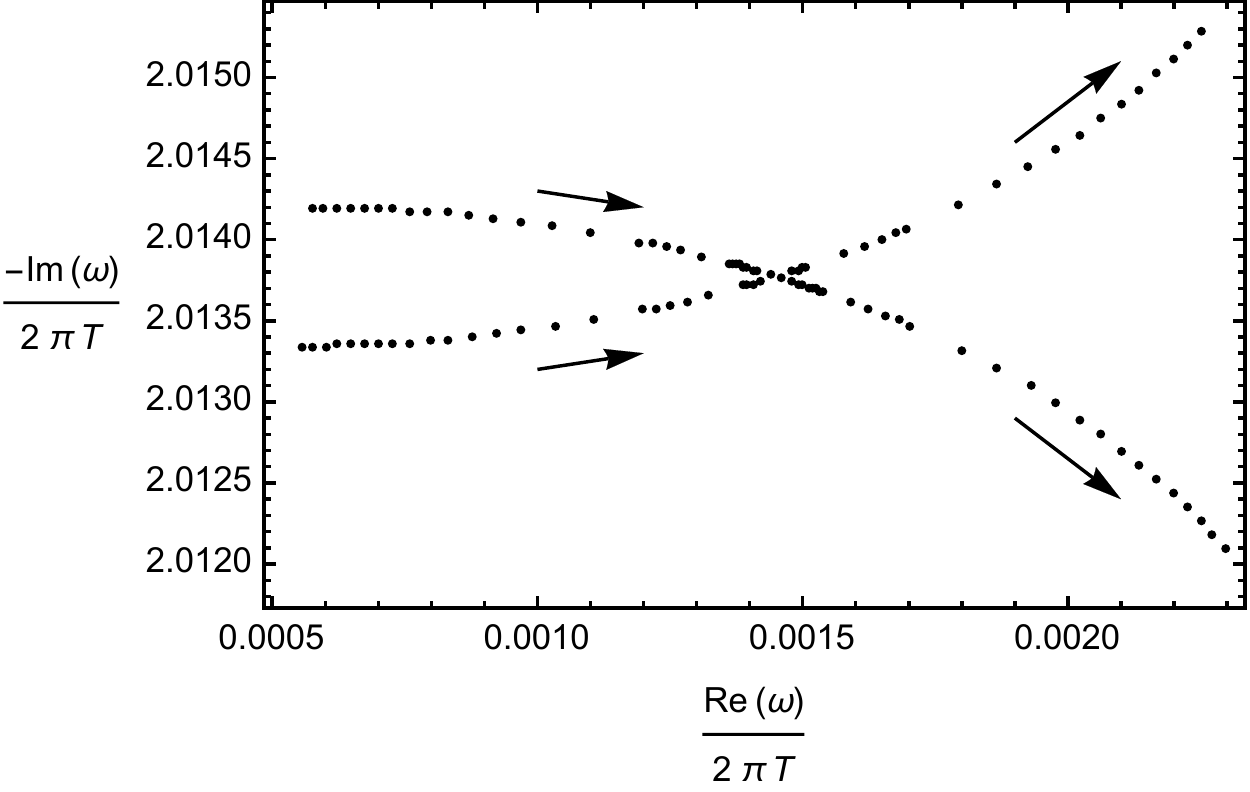}
\end{center}
\caption{Motion of the hydrodynamic (starting in the bottom left of the plot) and longest-lived infra red mode (starting in the top left of the plot) in the complex $\omega$ plane as $\left|k\right|/T$ is increased (from approximately $101.09$ to approximately $101.15$) at fixed $T/m=10^{-3}$ and fixed phase of the wavenumber $\phi_k=7.095\times 10^{-4}$. There is a collision for $\left|k\right|\simeq 101.12T$. {Equation \eqref{eq:Gnearcollision} predicts a collision at $\left|k\right|\simeq 101.125T$ and $\phi_k=7.374\times 10^{-4}$. In Figure \ref{fig:axionequilibrationdata} in Appendix \ref{sec:AxionAppendix}, we show that the discrepancy between the numerical and analytical values from equation \eqref{eq:Gnearcollision} decreases with temperature.}}
\label{fig:axioncollisionphase}
\end{figure}

In order to extract the existence of the pole collision, a more refined perturbative computation of $G_{\varepsilon\varepsilon}$ at the points $\omega=\omega_n+\delta\omega,k^2=k_n^2+\delta (k^2)$ is required. This yields
\begin{equation}
\label{eq:Gnearcollision}
G_{\varepsilon\varepsilon}^{-1}\left(\omega,k\right)\propto\left(\mathcal{D}_n\delta(k^2)-i\delta\omega\right)\left(1-i\tau_n\delta\omega\right)-i\lambda_n\delta\omega,
\end{equation}
where we show only terms relevant for understanding the collision. The low $T$ limit of each coefficient is
\begin{equation}
\begin{aligned}
\label{eq:taundefn}
\tau_n\rightarrow\frac{9m}{16\sqrt{6}(2+n)\pi^2T^2},\quad
\lambda_n\rightarrow\sqrt{\frac{3}{2}}(n(n+4)+3)\frac{\pi T}{m},
\end{aligned}
\end{equation}
while $\mathcal{D}_n\rightarrow D_\varepsilon$ in the same limit. The comparable size of the $\delta\omega$ and $\tau_n(\delta\omega)^2$ terms at frequencies $\delta\omega \sim T^2$ indicates that for such frequencies $G_{\varepsilon\varepsilon}$ is dominated by two poles, whose dispersion relations are given by solving the quadratic equation \eqref{eq:Gnearcollision} for $\delta\omega(\delta k)$. In Figure \ref{fig:axioncollision} we show that indeed \eqref{eq:Gnearcollision} correctly describes the locations of the two poles near $\omega_0$ {for real values of $k$, including the absence of a collision.} The poles collide (coincide in Fourier space) at the complex value of $\delta k$ where the discriminant of the quadratic polynomial vanishes, and the dispersion relation has a branch point. {This collision is shown in Figure \ref{fig:axioncollisionphase}.}

The collision closest to the origin of $k$-space ($n=0$) signals the breakdown of hydrodynamics. The absolute value of $k$ and $\omega$ at this collision are (as $T\rightarrow0$)
\begin{equation}
\begin{aligned}
\label{eq:axioncollisionscales}
\omega_{eq}&\,\equiv\left|\omega_{\text{collision}}\right|\rightarrow4\pi T\left(1+\frac{8\sqrt{6}\pi T}{9m}+\ldots\right),\\
k_{eq}^2&\,\equiv\left|k_{\text{collision}}\right|^2\rightarrow\frac{\omega_{eq}}{D_{\varepsilon}}\left(1-\frac{4\sqrt{6}\pi T}{3m}+\ldots\right),
\end{aligned}
\end{equation}
from which our main results \eqref{eq:relaxationtimeintro} and \eqref{eq:diffusivityintro} follow. 

The collision location asymptotically approaches real (imaginary) values of $k$ $(\omega)$ as $T\rightarrow0$. More precisely, as $T\rightarrow0$ the phases of $k$ and $\omega$ at the collision point are
\begin{equation}
\label{eq:axioncollisionphase}
\phi_{k}\rightarrow\frac{2^4}{6^{3/4}}\left(\frac{\pi T}{m}\right)^{3/2},\quad\quad \phi_{\omega}\rightarrow-\frac{\pi}{2}+\phi_{k},
\end{equation}
where $k_{\text{collision}}= k_{eq}e^{i\phi_k}$ and $\omega_{\text{collision}}=\omega_{eq}e^{i\phi_\omega}$.

\begin{figure}
\begin{center}
\includegraphics[width=0.7\textwidth]{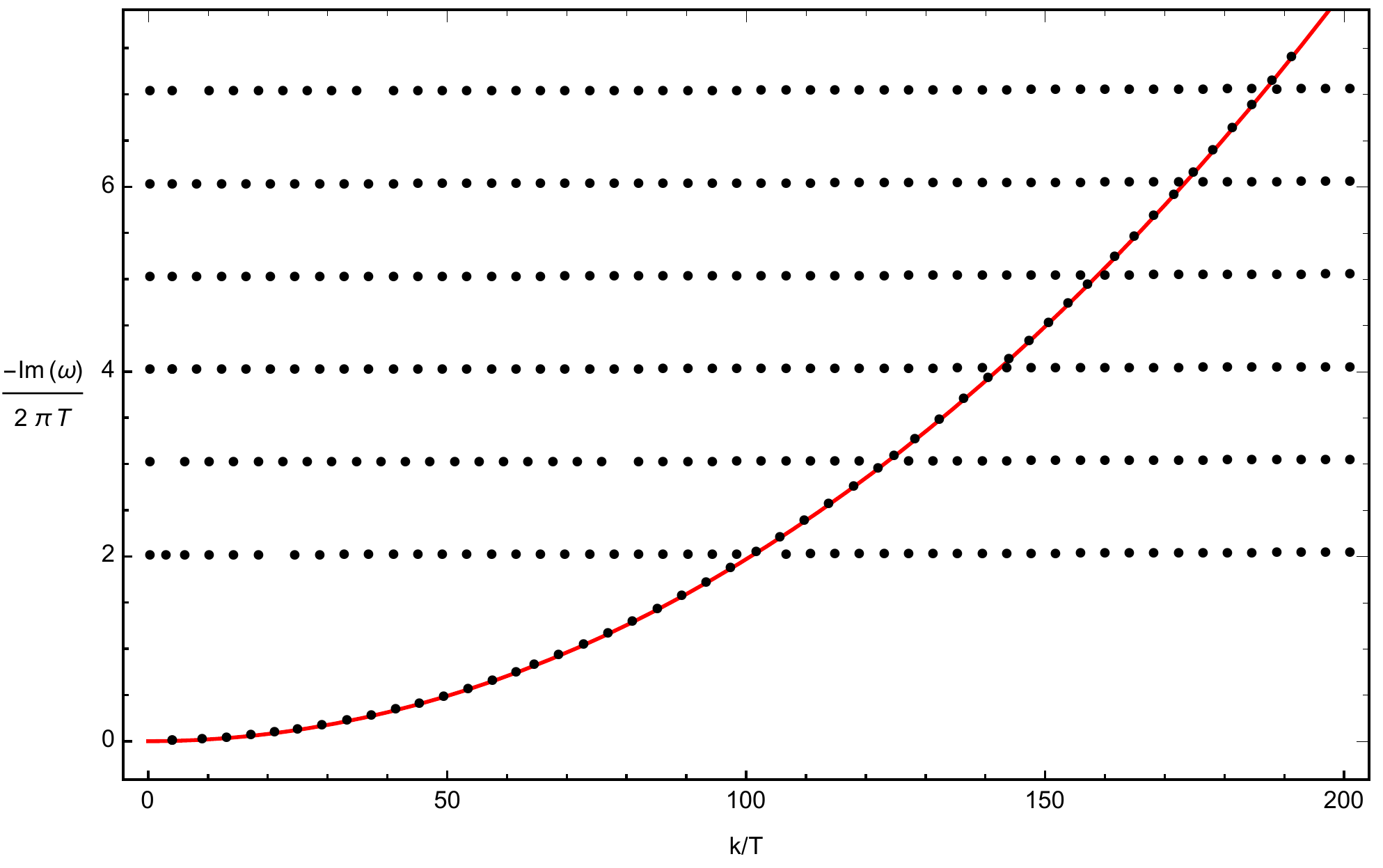}
\end{center}
\caption{Numerical results for the frequencies of the hydrodynamic and longest lived infra-red modes of the neutral, translation-symmetry breaking model at $T/m=10^{-3}$ (circles). Away from $\omega_{n\ge0}$, the analytic dispersion relation \eqref{eq:scalingdispersion} (solid red line) provides an excellent approximation to the exact location of a pole.}
\label{fig:axionoverviewallmodes}
\end{figure}

Note that even after the pole collision formally indicating the breakdown of hydrodynamics, {Figure \ref{fig:axionoverviewallmodes} }illustrates that the system continues to exhibit a diffusion-like mode described extremely well by the dispersion relation \eqref{eq:scalingdispersion}. {In the limit of zero temperature, the tower of infra-red poles in AdS$_2\times$R$^2$ coalesces in a branch cut along the imaginary axis \cite{Faulkner:2009wj,Edalati:2010hk}. As the branch cut passes through $k=0$, we expect that a hydrodynamic-like series for the dispersion relation at $T=0$ would contain non-analytic terms, as was found recently in \cite{Moitra:2020dal} for a similar state.} 

\section{Diffusion in a charged holographic state}

The AdS$_4$-RN solution to Einstein-Maxwell gravity 
\begin{equation}
S=\int d^4x\sqrt{-g}\left(R+6-\frac{1}{4}F^2\right),
\end{equation}
has a metric of the form \eqref{eq:planarmetric} but supported by a radial electric field
\begin{equation}
A_t=\mu\left(1-\frac{r_0}{r}\right),
\end{equation}
such that the emblackening factor is
\begin{equation}
\label{eq:RNemblackeningfunc}
f(r)=1-\left(1+\frac{\mu^2}{4r_0^2}\right)\frac{r_0^3}{r^3}+\frac{\mu^2r_0^2}{4r^4}.
\end{equation}
This solution represents a translationally invariant state with $U(1)$ chemical potential $\mu$, and its linear perturbations can be written in terms of four decoupled variables. Further details of the solution and the calculations underlying our results are given in Appendix \ref{sec:RNAppendix}.

The state exhibits two independent diffusive hydrodynamic modes, each associated to a different such variable. The first, corresponding to diffusion of energy and $U(1)$ charge with diffusivity $D_\varepsilon$, is analogous to the diffusive mode of the previous section.\footnote{\baselineskip1pt In the low $T$ limit both such modes have $D_\varepsilon=\kappa/c_\rho$ with $\kappa$ the open circuit thermal conductivity and $c_\rho$ the heat capacity \cite{Blake:2016jnn,Davison:2018nxm}.} The second corresponds to the transverse diffusion of momentum with diffusivity $D_\Pi$. In the limit of low temperature, the diffusivities are\footnote{See Appendix \ref{sec:RNAppendix} for the full{, $T$-dependent} expressions.}
\begin{equation}
{D_\varepsilon(T=0)=\frac{\sqrt3}{\mu}\,,\quad\quad\quad D_\Pi(T=0)=\frac1{\sqrt{12}\mu}\,.}
\end{equation}

The variables exhibiting each of these modes have $k\rightarrow0$ infra-red scaling dimensions \cite{Edalati:2009bi,Edalati:2010hk,Edalati:2010pn}
\begin{equation}
\label{eq:RNirscalingdimensions}
\Delta_{\varepsilon}(0)=2,\quad\quad\quad \Delta_{\Pi}(0)=1,
\end{equation}
and numerical calculations confirm that at small $k$ and $T$ each corresponding retarded Green's function exhibits non-hydrodynamic poles at the locations \eqref{eq:IRpolelocations}. As before, a collision close to the imaginary $\omega$ axis between the longest-lived such pole and the hydrodynamic pole signifies the independent breakdown of hydrodynamics in each case.

\begin{figure}
\begin{center}
\includegraphics[width=0.49\textwidth]{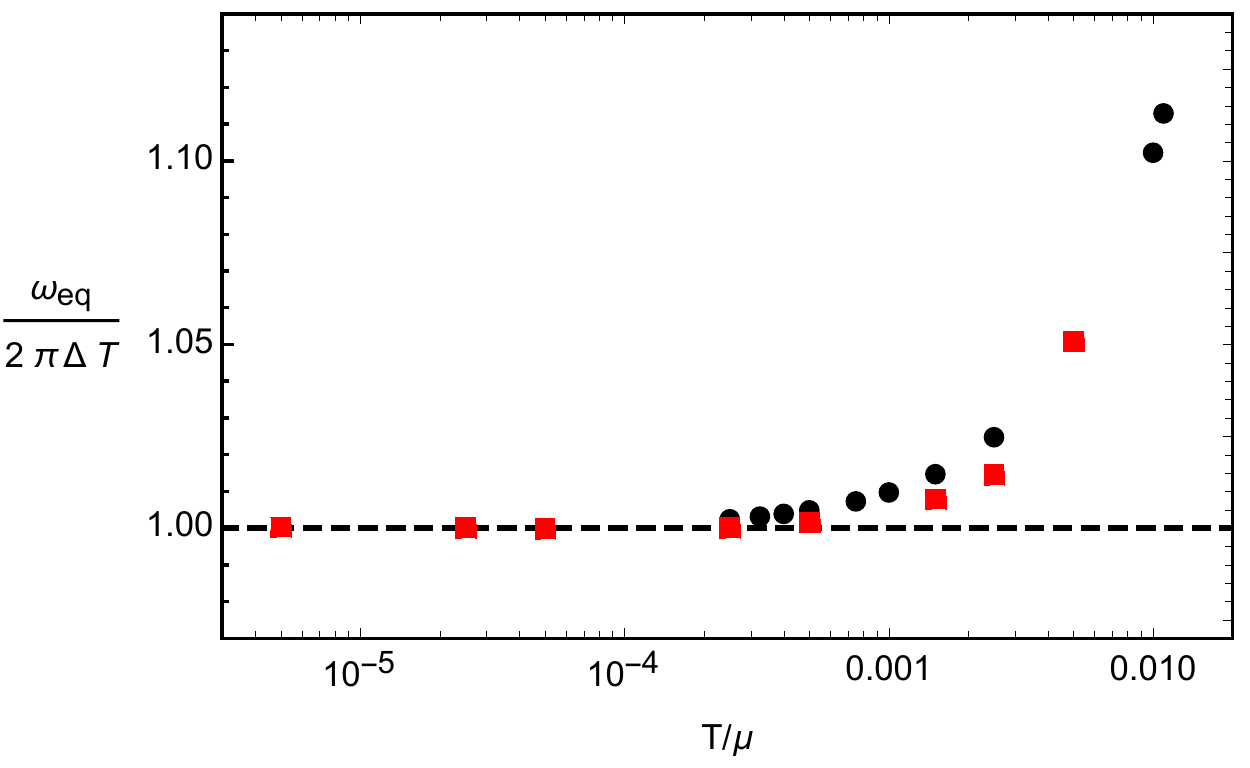}
\includegraphics[width=0.49\textwidth]{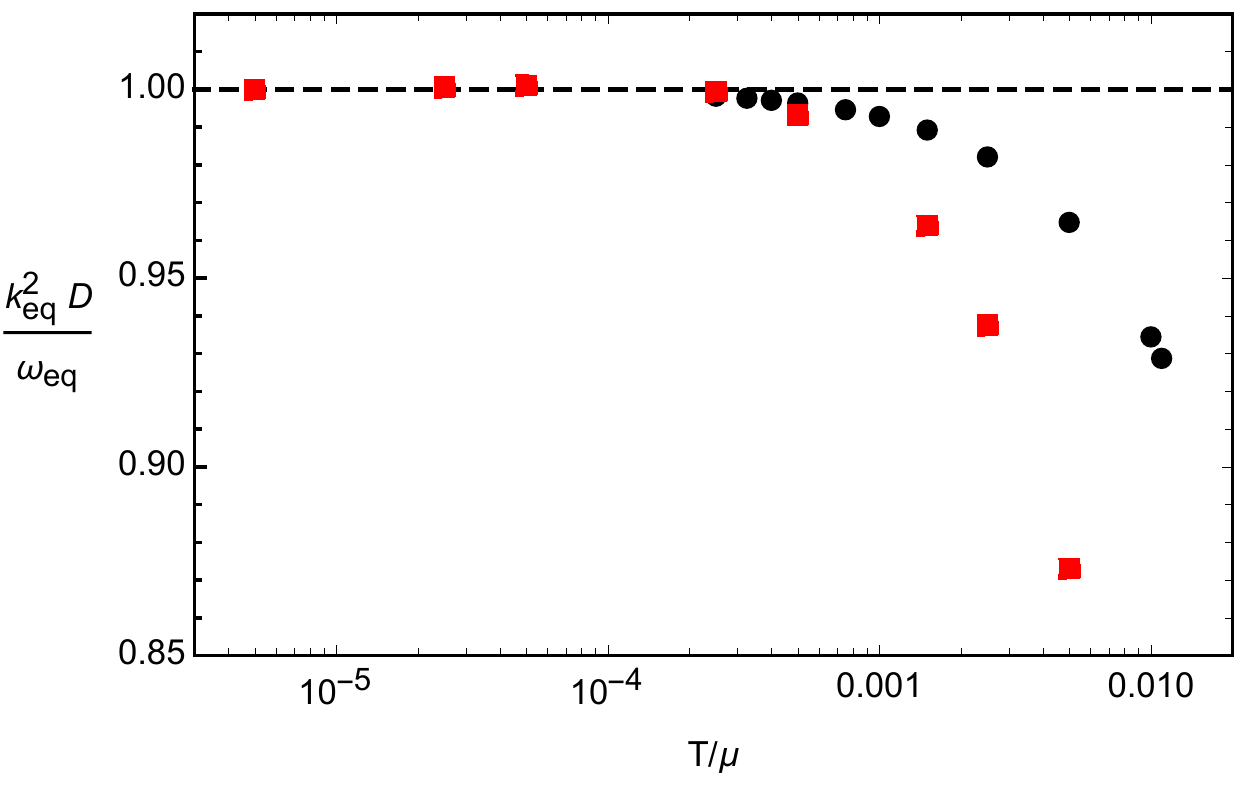}
\end{center}
\caption{Numerically obtained local equilibration data for diffusive hydrodynamics in $G_{\varepsilon\varepsilon}$ (black circles) and $G_{\Pi\Pi}$ (red squares) of the charged state.}
\label{fig:RNeqscalesmaintext}
\end{figure}

At low $T$ (and until the collision occurs) both hydrodynamic modes are described extremely well by the quadratic approximation to the hydrodynamic dispersion relation, while the locations of the longest-lived non-hydrodynamic poles depend only very weakly on $k$. As a consequence, the equilibration scales in both cases are set by the simple formulae \eqref{eq:relaxationtimeintro} and \eqref{eq:diffusivityintro} as shown in Figure \ref{fig:RNeqscalesmaintext}.

\begin{figure}
\begin{center}
\includegraphics[width=0.7\textwidth]{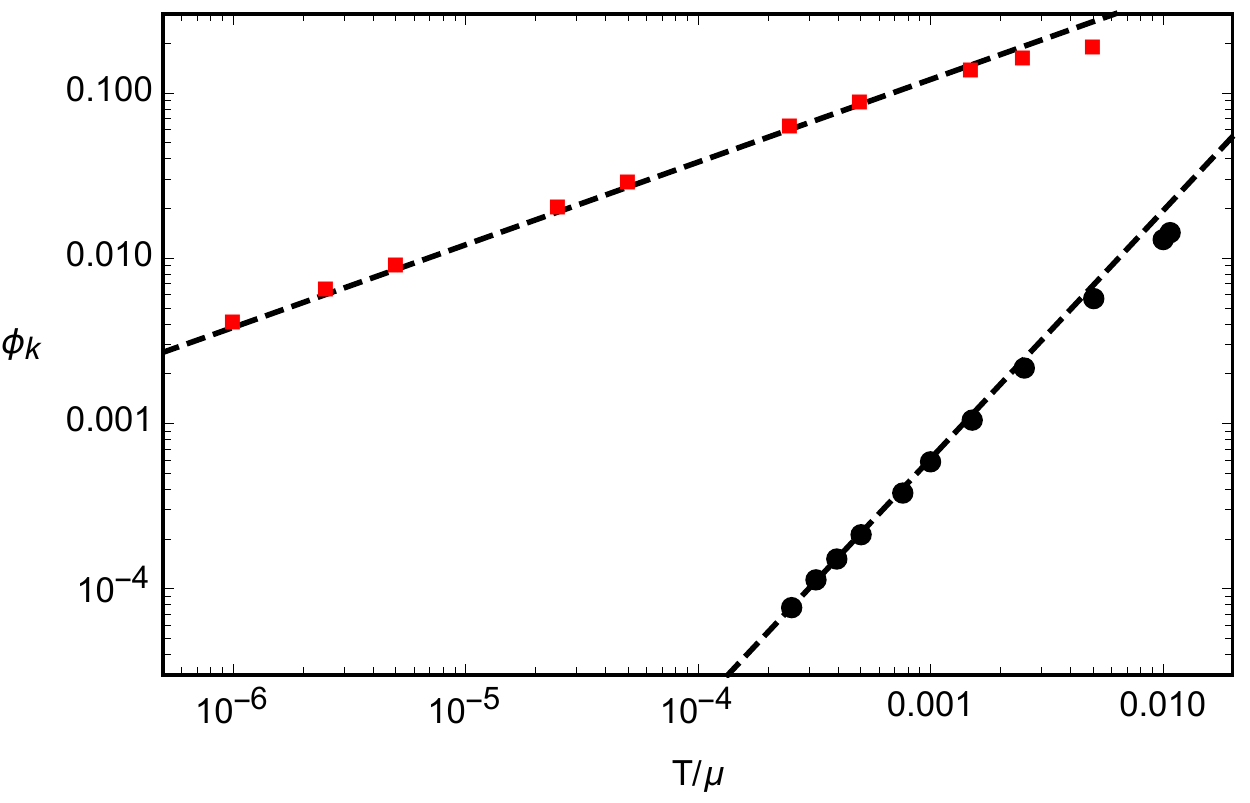}
\end{center}
\caption{Numerically obtained $\phi_{k}$ for $G_{\varepsilon\varepsilon}$ (black circles) and $G_{\Pi\Pi}$ (red squares) of the charged state. Dashed lines shows the best-fits to a power law at small $T$: $\phi_k=19.2(T/\mu)^{1.50}$ and $\phi_k=3.80 (T/\mu)^{0.50}$ respectively.}
\label{fig:RNphasemaintext}
\end{figure}

The phase of the collision wavenumber $\phi_{k}$ in each case is shown in Figure \ref{fig:RNphasemaintext}, illustrating that the collision point asymptotically approaches real values of $k$ as $T\rightarrow0$. Both cases here, and the result \eqref{eq:axioncollisionphase} in the previous example, are consistent with $\Delta$ controlling the low $T$ scaling of the phase via $\phi_{k}\sim T^{\Delta-1/2}$.

As in the previous example, the system continues to exhibit diffusion-like modes even after the collision formally indicating the breakdown of hydrodynamics. The dispersion relations of these modes are extremely well approximated by the quadratic approximation to diffusive hydrodynamics, as shown in Figure \ref{fig:RNallmodes} of Appendix \ref{sec:RNAppendix}.

\section{Comparison with SYK chain}

The SYK model is a (0+1)-dimensional theory of $N$ interacting fermions that, in the limit of large $N$ and strong interactions, is governed by the same effective action as a theory of gravity in a nearly-AdS$_2$ spacetime \cite{SY92,Sachdev:2010um,Almheiri:2014cka,Kitaevtalk,Sachdev:2015efa,Maldacena:2016hyu,Maldacena:2016upp,Jensen:2016pah,Engelsoy:2016xyb}. The SYK chain \cite{Gu:2016oyy} is a higher-dimensional generalisation of this, which has served as a very useful toy model for studying diffusive energy transport in strange metal states of matter. As it exhibits the local quantum criticality characteristic of AdS$_2\times $R$^2$ fixed points, it is natural to ask whether local equilibration in this explicit microscopic model is governed by our general results \eqref{eq:relaxationtimeintro} and \eqref{eq:diffusivityintro}. 

In \cite{Choi:2020tdj}, an SYK chain model with $N$ Majorana fermions per site $\chi_{i,x}$ and Hamiltonian 
\begin{equation}
\begin{aligned}
H=&\,i^{q/2}\sum_{x=0}^{M-1}\left(\sum_{1\leq i_1<\ldots< i_q\leq N}J_{i_1\ldots i_q,x}\chi_{i_1,x}\ldots\chi_{i_q,x}\right.\\
&\,+\left.\sum_{\substack{1\leq i_1<\ldots<i_{q/2}\leq N \\ 1\leq j_1<\ldots<j_{q/2}\leq N}}J'_{i_1\ldots i_{q/2}j_1\ldots j_{q/2},x}\chi_{{i_1},x}\ldots\chi_{i_{q/2},x}\chi_{j_1,x+1}\ldots\chi_{j_{q/2},x+1}\right),
\end{aligned}
\end{equation}
was studied. The two terms represent $q$-body on-site and nearest neighbour interactions respectively, where the couplings $J_{i_1\ldots i_q,x}$ and $J'_{i_1\ldots i_{q/2}j_1\ldots j_{q/2},x}$ are Gaussian random variables with zero mean. Remarkably, in the limit $N\gg q^2\gg1$ an exact analytic expression for $G_{\varepsilon\varepsilon}$ was found for all values of the effective interaction strength $0<v<1$ and the relative strength of on-site and inter-site interactions $0<\gamma\leq1$ \cite{Choi:2020tdj}. The analytic expression is given in Appendix \ref{sec:SYKAppendix}, where the details of the model are also summarised. 

\begin{figure}
\begin{center}
\includegraphics[width=0.49\textwidth]{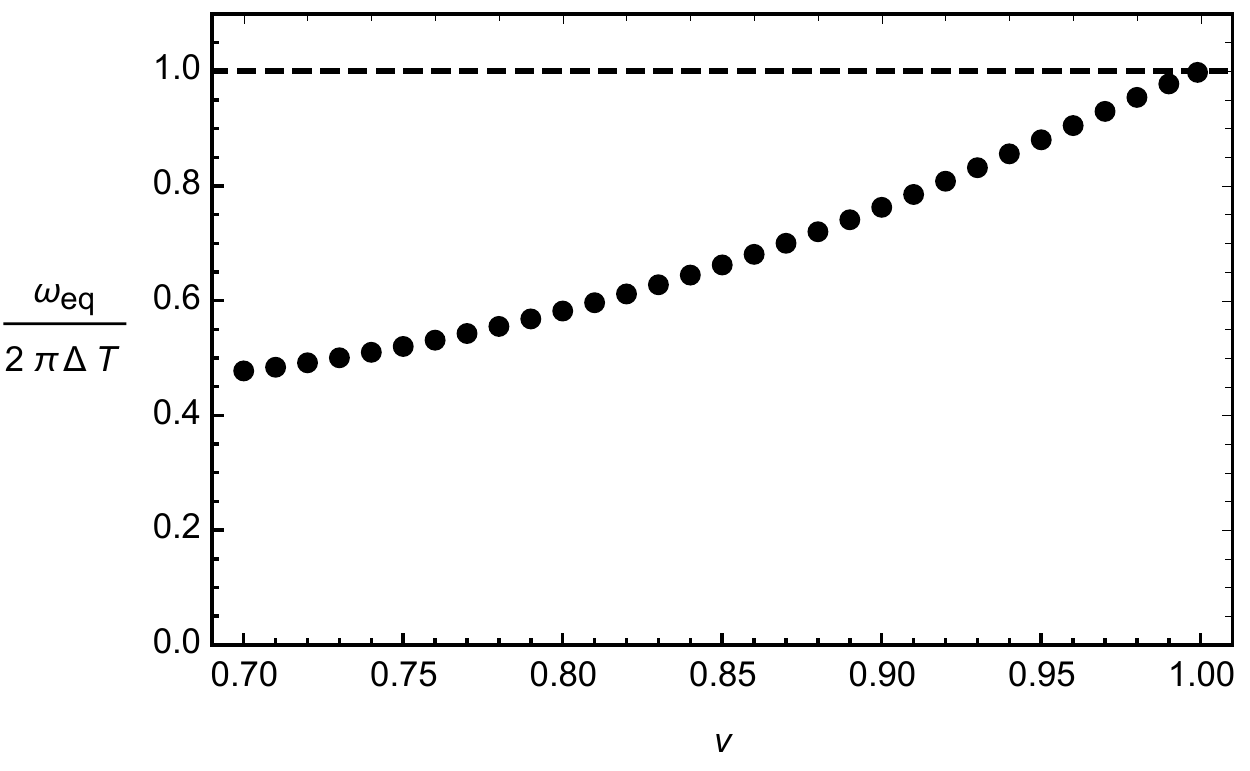}
\includegraphics[width=0.49\textwidth]{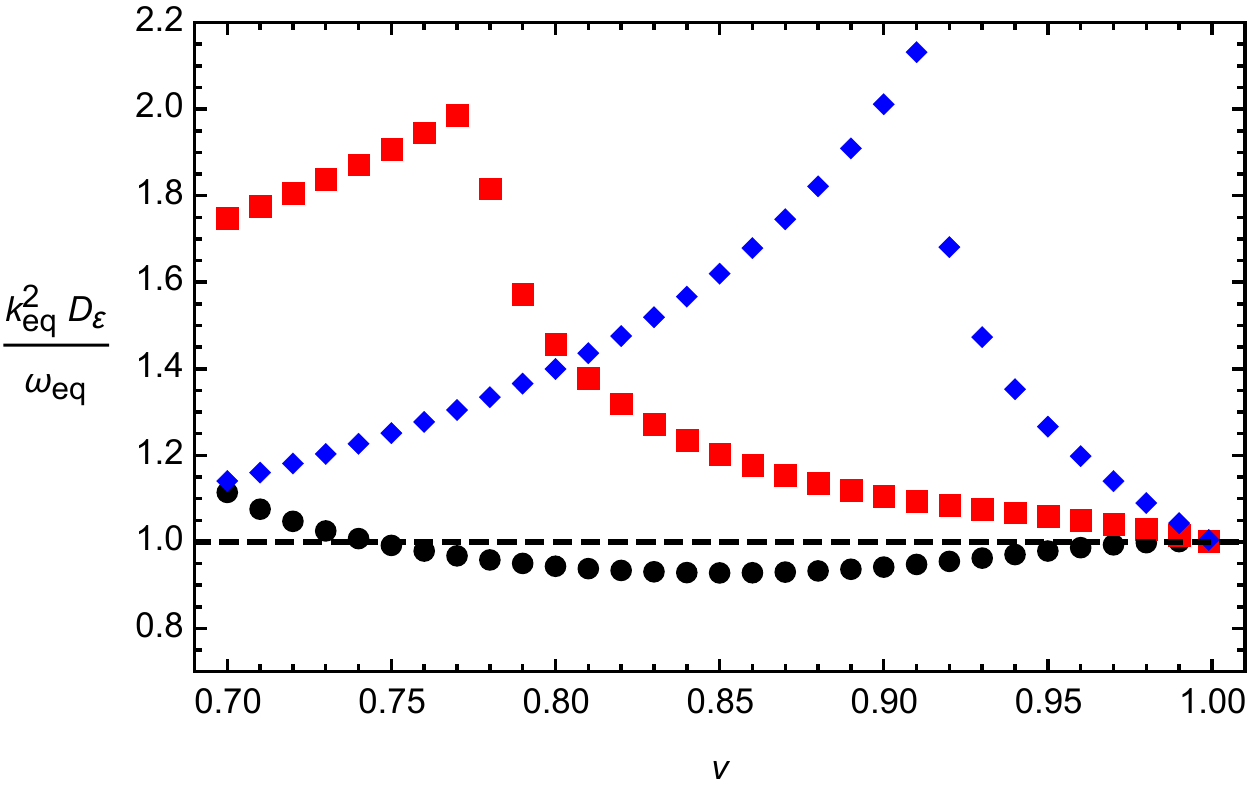}
\end{center}
\caption{Local equilibration frequency (left panel) and diffusivity ratio (right panel) for the large-$q$ SYK chain model. $\omega_{eq}$ is $\gamma$-independent, while the diffusivity ratio is shown for $\gamma=1$ (black circles), $\gamma=0.4$ (red squares) and $\gamma=0.2$ (blue diamonds).}
\label{fig:SYKplotmaintext}
\end{figure}

Taking advantage of this result, a detailed study of the breakdown of diffusive hydrodynamics as a function of interaction strength $v$ was performed in \cite{Choi:2020tdj}. Here we will focus on the limit of strong interactions $v\rightarrow1${, which is equivalent to $T\to0$}. In this limit, the longest-lived non-hydrodynamic modes are a series of infra-red modes located at precisely the frequencies $\omega_n$ of equation \eqref{eq:IRpolelocations} with $\Delta=2$ as $k\rightarrow0$. Hydrodynamics breaks down due to a collision between the hydrodynamic mode and the longest-lived of these infra-red modes, and in Figure \ref{fig:SYKplotmaintext} we confirm that the local equilibration scales are given simply by
\begin{equation}
\omega_{eq}\rightarrow2\pi \Delta T\quad\quad\text{and}\quad\quad k_{eq}^2\rightarrow\frac{\omega_{eq}}{D_\varepsilon}\quad\quad\quad\text{as}\quad\quad v\rightarrow1,
\end{equation}
analogously to \eqref{eq:relaxationtimeintro} and \eqref{eq:diffusivityintro}.
Unlike in the holographic examples, the pole collision at strong interactions here happens for real $k$ (i.e. $\phi_k=0$). It would be interesting to determine whether finite $1/q$ corrections generate a small non-zero phase $\phi_k$. Consistently with the other examples we have presented, following the formal breakdown of hydrodynamics the spectrum still contains a mode whose dispersion relation is very well approximated by the quadratic approximation to diffusive hydrodynamics.

\section{Outlook}

There is good reason to expect that at least some of our results will generalise beyond the specific examples studied here to other states governed by AdS$_2$ infra-red fixed points. While our key observation that the quadratic approximation to the hydrodynamic dispersion relation works parametrically well even for wavenumbers $k^2\sim T$ seems unusual, it is non-trivially consistent with the result \eqref{eq:Dchaosintro} that is indeed true for holographic AdS$_2\times$R$^2$ fixed points with a universal deformation \cite{Blake:2016jnn} as well as in related SYK chain models \cite{Gu:2016oyy,Davison:2016ngz}. 

There are more general holographic and SYK-like systems governed by AdS$_2$ fixed points that exhibit additional diffusive modes beyond the two types we have studied. Of particular interest are non-translationally invariant systems with a $U(1)$ symmetry, for which an Einstein relation relates the electrical resistivity to a diffusivity \cite{Hartnoll:2014lpa,Blake:2016jnn,Davison:2016ngz}. If our results \eqref{eq:relaxationtimeintro} and \eqref{eq:diffusivityintro} extend to such modes, they will therefore also provide a simple relation between the phenomenologically important electrical resistivity and the local equilibration scales of such strongly correlated systems.

Confirmation of the broader applicability of our result \eqref{eq:Dscalesintro} for AdS$_2\times$R$^2$ solutions would be an important step for quantifying diffusivities near general infra-red fixed points. One way to do this would be to identify a speed $u$ and timescale $\tau$ such that in general $D\sim u^2\tau$ with the coefficient being $T$-independent. This is difficult even for the relatively simple case of holographic energy diffusion, primarily because there are two exceptional types of fixed point where dangerously irrelevant deformations take over the properties of the mode: AdS$_2\times$R$^2$ fixed points (i.e.~dynamical critical exponent $z=\infty$) \cite{Blake:2016jnn} and relativistic fixed points (i.e.~$z=1$) \cite{Blake:2017qgd}. If our result does generalize to AdS$_2\times$R$^2$ solutions (including those with non-universal deformations), both of these exceptional cases will be consistent with the identification $u=v_{eq}$ and $\tau=\tau_{eq}$.\footnote{\baselineskip1pt We expect the result of footnote \ref{footnoteslowmode} to apply to the $z=1$ cases due to the existence of a parametrically slow mode \cite{Davison:2018ofp,Davison:2018nxm}.}${}^{,}$\footnote{\baselineskip1pt Another family of holographic $z=1$ fixed points was studied in \cite{Betzios:2017dol,Betzios:2018kwn}. These geometries do not have a slow mode, but in a certain limit also display an emergent SL($2$,R) symmetry and a spectrum of infra-red mode similar to the one studied in this work.} Provided that naive $T$-scaling holds for the equilibration scales in the other cases {with finite Lifshitz exponent $z$} ($\tau_{eq}\sim 1/T$ and $v_{eq}\sim T^{1-1/z}$), which seems likely, this identification will then work for all fixed points.

For cases where the breakdown of hydrodynamics is due to a slow mode, it is the separation of scales between the decay rate of the slow mode $\Gamma$ and that of typical non-hydrodynamic excitations $T$ that allows one to augment the hydrodynamic description to incorporate the slow mode. Mathematically, the separation allows one to resum the hydrodynamic expansion into a square root form, valid at scales $\omega\sim k\sim\Gamma$ (see e.g.~\cite{Davison:2014lua,Chen:2017dsy,Grozdanov:2018fic}). This makes manifest the convergence properties of the hydrodynamic expansion, $k_{eq}\sim\Gamma$. It would be very interesting if one could extract an analogous effective theory for the cases described here, taking advantage of the separation of scales between the decay rate of the infra-red modes (set by $T$) and that of the other non-hydrodynamic excitations (set by the curvature of AdS$_2$). Such an effective theory would need to resum the effects of the entire tower of infra-red modes and would give a greater understanding of why the quadratic approximation to diffusive hydrodynamics is valid up to (and indeed beyond) the wavenumber $k_{eq}$.

It would also be interesting to study whether our results continue to hold for AdS$_2$ fixed points supported by a different hierarchy of scales (such as large angular momentum or magnetic field compared to temperature), and whether analogous results hold for other types of hydrodynamic modes near these fixed points.

It would be very useful to have a semi-holographic description of our results (along the lines of \cite{Faulkner:2010tq,Nickel:2010pr}), in which we couple a gapless hydrodynamic diffusion mode to the tower of infra-red modes associated with the AdS$_2$ region of the spacetime. Our results rely on the fact that the the two types of mode have very little effect on one another (see e.g.~Figure \ref{fig:axion2modesmaintext}) and a semi-holographic description may clarify exactly under what conditions this is the case. A related avenue would be to adapt the recently constructed holographic effective Schwinger-Keldysh action for diffusion to AdS$_2$ horizons \cite{Nickel:2010pr,deBoer:2015ija,deBoer:2018qqm,Glorioso:2018mmw}.

{Extending our work to Schrödinger $z=2$ IR geometries would allow to make contact with ultracold atomic systems \cite{Son2008}, which realize a strongly-interacting Fermi gas near unitarity \cite{Enss2019}. 

In \cite{Brown_2018}, the charge diffusivity of a cold atomic system coupled to an optical lattice was measured. In \cite{Zhang:2016ofh,Zhang_2019}, the thermal diffusivity of high $T_c$ superconductors in the strange metallic regime was also reported. It would be interesting to investigate the deviations from diffusive hydrodynamics in these systems.}

The examples we studied are all consistent with bounds on $D$ of the type proposed in \cite{Hartman:2017hhp,Lucas:2017ibu} but with the velocity in the bound given by $v_{eq}$ (rather than the operator growth velocity, characteristic velocity of low energy excitations, or butterfly velocity).\footnote{\baselineskip1pt While the SYK chain results shown in Figure \ref{fig:SYKplotmaintext} do not obey a strict bound $D\leq v_{eq}^2\tau_{eq}$, there is no parametric violation of such a relation and thus they are consistent with \cite{Hartman:2017hhp,Lucas:2017ibu}.} Indeed, the arguments in \cite{Hartman:2017hhp,Lucas:2017ibu} assume that $v_{eq}$ is set by one of these velocities and thus imply the bound $D\lesssim v_{eq}^2\tau_{eq}$. In this sense our results support the assumption of \cite{Hartman:2017hhp,Lucas:2017ibu} that the local equilibration is controlled by an underlying effective lightcone, even though the systems are non-relativistic. It would be very worthwhile to determine $D$, $v_{eq}$ and $\tau_{eq}$ for other states, and over a wider parameter range, in order to establish the robustness of these observations and to determine whether $v_{eq}$ is set by a speed such as the butterfly velocity in general.

\begin{acknowledgments}

We are grateful to Changha Choi, Luca Delacr\'etaz, Sa\v{s}o Grozdanov, Sean Hartnoll, M\'{a}rk Mezei, Subir Sachdev, G\'{a}bor S\'{a}rosi, and Benjamin Withers for helpful discussions. 
D.~A. is supported by the `Atracci\'on de Talento' programme (2017-T1/TIC-5258, Comunidad de Madrid) and through the grants SEV-2016-0597 and PGC2018-095976-B-C21.
The work of R.~D. is supported by the STFC Ernest Rutherford Grant ST/R004455/1. The work of B.~G. is supported by the European Research Council (ERC) under the European Union's Horizon 2020 research and innovation program (grant agreement No758759). 
The work of K.~S. is supported by the European Research Council (ERC) under the European Union's Horizon 2020 research and innovation program (grant agreement No758759 and No818066).

 \end{acknowledgments}
 
 \bibliography{univIRrelax-biblio}
 
 \pagebreak

\appendix

\section{Perturbations in the neutral translation-breaking model}
\label{sec:AxionAppendix}

In this appendix we present in more detail the calculations leading to the results presented in the main text for the neutral translation-breaking model of \cite{Bardoux:2012aw,Andrade:2013gsa}. The action of this model is
\begin{equation}
S=\int d^4x\sqrt{-g}\left(R+6-\frac{1}{2}\sum_{i=1}^{2}\left(\partial\varphi_i\right)^2\right),
\end{equation}
and it has the classical solution with metric
\begin{equation}
\begin{aligned}
\label{eq:adsaxionappendix}
ds^2=-r^2f(r)dt^2+r^2\underline{dx}^2+\frac{dr^2}{r^2f(r)},\quad\quad\quad f(r)=1-\frac{m^2}{2r^2}-\left(1-\frac{m^2}{2r_0^2}\right)\frac{r_0^3}{r^3},
\end{aligned}
\end{equation}
and scalar field profiles $\varphi_i=mx^i$ $(i=1,2)$ that break translational symmetry. The horizon is located at $r_0$ and the asymptotically AdS boundary at $r\rightarrow\infty$. The Hawking temperature of the solution is
\begin{equation}
\label{eq:axionT}
T=\frac{3r_0}{4\pi}\left(1-\frac{m^2}{6r_0^2}\right),
\end{equation}
and further details of its thermodynamic properties can be found in \cite{Andrade:2013gsa}. 

At zero temperature, $m^2=6r_0^2$ and an AdS$_2\times$R$^2$ metric emerges near the extremal horizon. To see this explicitly, one should change coordinates from $(t,r)$ to $(u,\zeta)$ where
\begin{equation}
\label{eq:axionAdS2coords}
r=r_0+\epsilon\,\zeta,\quad\quad\quad t=\frac{u}{\epsilon},
\end{equation}
and then take the $\epsilon\rightarrow0$ limit of \eqref{eq:adsaxionappendix} to obtain
\begin{equation}
ds^2\rightarrow-\frac{\zeta^2}{L^2}du^2+L^2\frac{d\zeta^2}{\zeta^2}+r_0^2\;\underline{dx}^2,
\end{equation}
where the AdS$_2$ radius of curvature is $L^2=1/3$.

The hydrodynamics of this model were studied and explained in \cite{Davison:2014lua}. Over the longest distance and time scales, this system supports a single hydrodynamic mode corresponding to the diffusion of energy. The dispersion relation of this mode is
\begin{equation}
\begin{aligned}
\label{eq:axionexactD}
\omega_{\text{hydro}}(k)=-iD_\varepsilon k^2+O(k^4),\quad\quad\quad D_\varepsilon=\frac{1}{m^2}\sqrt{\frac{3}{2}m^2+4\pi^2T^2}.
\end{aligned}
\end{equation}

\subsection{Gauge-invariant perturbations}

Perturbations of this spacetime can be conveniently studied by defining suitable gauge-invariant combinations of the Fourier space perturbations of the metric components and matter fields. Specifically, there is a choice of four such variables for which the equations decouple. The decoupling is a reflection of the diagonalisation of the matrix of Green's functions of the dual operators. Of these variables, the one relevant for the energy density Green's function is \cite{Davison:2014lua}
\begin{equation}
\begin{aligned}
\label{eq:linaxmastervble}
\tilde{\psi}(r,\omega,k)=&\;\frac{r^4 f}{(k^2+r^3f')} \left[ \frac{d}{dr}\left(\frac{\delta g_{xx}+\delta g_{yy}}{r^2}\right) - \frac{2i k}{r^2}\delta g_{xr}- 2rf\delta g_{rr} - \frac{k^2+r^3 f'}{r^5 f}\delta g_{yy} \right] \\
&\;-\frac{mr}{2\left(k^2+m^2\right)} \Big( \frac{m}{r^2}(\delta g_{xx}-\delta g_{yy}) - 2i k \delta\varphi_1\Big),
\end{aligned}
\end{equation}
where $\omega$ and $k$ here are the frequency and wavenumber after a Fourier transform with respect to the $t$ and $x^1$ coordinates in which the metric has the form \eqref{eq:adsaxionappendix}. This variable obeys the equation
\begin{equation}
\label{eq:axionGIeq1}
\left(r^2 f \tilde{\psi}' \right)' + \left( \frac{\omega^2-k^2 f}{r^2f} + V(r) \right) \tilde{\psi} = 0,
\end{equation}
where primes denote derivatives with respect to $r$ and
\begin{equation}
\begin{aligned}
V(r)= -&\;\frac{3r_0\left(m^2-2r_0^2\right)}{2r^3\left(2k^2r+m^2\left(2r-3r_0\right)+6r_0^3\right)^2} \Big(4k^4 r^2 + m^4\left(-4r^2+6r_0r-3r_0^2\right)\\
		&\;+ 12m^2r_0\left(r^3-r_0^2r+r_0^3\right)-12r_0^3\left(2r^3+r_0^3\right) \Big).
\end{aligned}
\end{equation}

For some calculations it will be convenient for us to convert to an ingoing Eddington-Finkelstein like coordinate system with time coordinate $v=t+r_*$, where the tortoise coordinate $r_*$ is
\begin{equation}
r_*=\frac{1}{4\pi T}\left\{\log\left(\frac{r-r_0}{\sqrt{r^2+rr_0+r_0^2-\frac{m^2}{2}}}\right)+\frac{(m^2-3r_0^2)}{2r_0\sqrt{2m^2-3r_0^2}}\log\left(\frac{2r+r_0+\sqrt{2m^2-3r_0^2}}{2r+r_0-\sqrt{2m^2-3r_0^2}}\right)\right\}.
\end{equation}
In this coordinate system the relevant gauge-invariant variable may be written as
\begin{equation}
\begin{aligned}
\psi(r,\omega,k)=&\;r^4f\Biggl[\frac{d}{dr}\left(\frac{\delta g_{xx}+\delta g_{yy}}{r^2}\right)-\frac{i\omega}{r^4f}\left(\delta g_{xx}+\delta g_{yy}\right)-\frac{2ik}{r^2}\left(\delta g_{xr}+\frac{\delta g_{vx}}{r^2f}\right)\\
&-2rf\left(\delta g_{rr}+\frac{2}{r^2f}\delta g_{vr}+\frac{1}{r^4f^2}\delta g_{vv}\right)-\frac{k^2+r^3f'}{r^5f}\delta g_{yy}\Biggr]\\
&-\frac{mr\left(k^2+r^3f'\right)}{2\left(k^2+m^2\right)}\left(\frac{m}{r^2}\left(\delta g_{xx}-\delta g_{yy}\right)-2ik\delta\varphi_1\right),
\end{aligned}
\end{equation}
where $\omega$ here denotes the frequency after a Fourier transform with respect to the $v$ coordinate. This variable obeys the equation of motion
\begin{equation}
\label{eq:axionEFeqn}
\frac{d}{dr}\left[\frac{r^2fe^{-2i\omega r_*}}{(k^2+r^3f')^2}\psi'\right]-\Omega(\omega,k)\frac{e^{-2i\omega r_*}}{r^2(k^2+r^3f')^3}\psi=0,
\end{equation}
where
\begin{equation}
\label{eq:Omegadefn}
\Omega(\omega,k)=(2r_0^2-m^2)3ir_0\omega+k^2(k^2+m^2).
\end{equation}
Up to contact terms, the energy density Green's function can be expressed as \cite{Blake:2018leo}
\begin{equation}
\label{eq:axionG}
G_{\varepsilon\varepsilon}(\omega,k)=k^2(k^2+m^2)\left.\frac{\psi}{-r^2\psi'+i\omega\psi}\right|_{r\rightarrow\infty},
\end{equation}
where in this expression $\psi$ denotes the solution of \eqref{eq:axionEFeqn} that is regular at the horizon $r=r_0$ in ingoing coordinates. We will fix the overall normalisation of the solution as $\psi(r_0,\omega,k)=1$, without loss of generality.

\subsection{Near-horizon perturbation equations}

We begin by analysing the relevant perturbation equation in the AdS$_2\times$R$^2$ spacetime that emerges near the horizon at low temperatures. To do this, we follow \cite{Faulkner:2009wj} in introducing the location of the extremal horizon $r_e$ via $6r_e^2=m^2$ and then defining
\begin{equation}
r=r_e+\epsilon\,\zeta,\quad\quad\quad r_0=r_e+\epsilon\,\zeta_0,\quad\quad\quad \omega=\epsilon\,\zeta_\omega.
\end{equation}
This is similar to the Fourier space version of the limit \eqref{eq:axionAdS2coords}, but now keeping $\omega/T$ fixed as $\omega\rightarrow 0$. The temperature is proportional to $\zeta_0$. At leading order in the near-horizon, low temperature limit (i.e.~$\epsilon\rightarrow0$ limit), the equation \eqref{eq:axionGIeq1} becomes
\begin{equation}
\label{eq:AdS2scalareqn}
\partial_\zeta^2\tilde{\psi} + \frac{2\zeta}{\zeta^2 - \zeta_0^2} \partial_\zeta\tilde{\psi}+\left(\frac{\zeta_\omega^2}{9\left(\zeta^2-\zeta_0^2\right)^2}-\frac{2\left(1+\frac{k^2}{m^2}\right)}{\zeta^2-\zeta_0^2}\right) \tilde{\psi}=0.
\end{equation}
This is the equation of a scalar field in non-zero temperature AdS$_2$ with effective mass $L^2m_{\text{eff}}^2=2\left(1+k^2/m^2\right)$. As such, we can associate it with an operator of dimension $\Delta(k)=(1+\sqrt{9+8k^2/m^2})/2$ at the infra-red fixed point of the field theory. The infra-red Green's functions $\mathcal{G}_{IR}$ of such operators can be found explicitly by solving the equation \eqref{eq:AdS2scalareqn} and imposing the usual AdS/CFT rules at the AdS$_2$ boundary $\zeta\rightarrow\infty$ \cite{Faulkner:2009wj,Hartnoll:2012rj}
\begin{equation}
\label{eq:appendixIRgreensfn}
\mathcal{G}_{IR}\propto T^{2\Delta(k)-1}\frac{\Gamma\left(\frac{1}{2}-\Delta(k)\right)\Gamma\left(\Delta(k)-\frac{i\omega}{2\pi T}\right)}{\Gamma\left(\frac{1}{2}+\Delta(k)\right)\Gamma\left(1-\Delta(k)-\frac{i\omega}{2\pi T}\right)}.
\end{equation}
At any non-zero temperature the infra-red Green's function has a series of poles along the negative imaginary frequency axis at the locations
\begin{equation}
\label{eq:appendixIRpolelocations}
\omega_n=-i2\pi T(n+\Delta(k)),\quad\quad\quad n=0,1,2,\ldots.
\end{equation}
To obtain $G_{\varepsilon\varepsilon}$, one must extend the near-horizon solutions through the rest of the spacetime. In practice this is very difficult to do, but schematically one expects a result of the form \cite{Faulkner:2009wj,Hartnoll:2012rj}
\begin{equation}
\label{eq:GIRmatched}
G_{\varepsilon\varepsilon}=\frac{A+B\mathcal{G}_{IR}+\ldots}{C+D\mathcal{G}_{IR}+\ldots},
\end{equation}
in the low frequency limit. If the expression written in \eqref{eq:GIRmatched} were exact (ignoring the ellipses), then at the locations \eqref{eq:appendixIRpolelocations} corresponding to poles of $\mathcal{G}_{IR}$ there would not be a pole of $G_{\varepsilon\varepsilon}$: the purported pole of $G_{\varepsilon\varepsilon}$ (arising from a pole of $\mathcal{G}_{IR}$ in the numerator) would cancel exactly with a zero (arising from a pole of $\mathcal{G}_{IR}$ in the denominator). However, the result \eqref{eq:GIRmatched} is not exact and numerical calculations show that while there are a pole and a zero of $G_{\varepsilon\varepsilon}$ at low $T$ near each location \eqref{eq:appendixIRpolelocations}, for any non-zero $T,k$ they are not exactly coincident. As a consequence, $G_{\varepsilon\varepsilon}$ does exhibit poles at the locations \eqref{eq:appendixIRpolelocations} at small $T$ and $k$ and we will present numerical results demonstrating this later in this Appendix. The qualitative observation that the emergence of an AdS$_2\times$R$^2$ spacetime at low temperatures coincides with the emergence of a series of poles of $G_{\varepsilon\varepsilon}$ with imaginary frequencies $\sim T$ has been made previously for other spacetimes (see e.g.~\cite{Edalati:2010hk,Edalati:2010pn}). We are going beyond this by providing the precise locations of these poles as $k,T\rightarrow0$.

\subsection{Low temperature dispersion relation}

In this section we will derive the dispersion relation \eqref{eq:scalingdispersion} for a pole of $G_{\varepsilon\varepsilon}$ in the limit of low $T$ (with $k^2/T$ fixed). The key observation that will allow us to make progress is that the equation \eqref{eq:axionEFeqn} simplifies considerably when $\Omega=0$ i.e.~when $\omega=ik^2(k^2+m^2)/(3r_0(2r_0^2-m^2))$, allowing us to formally obtain $G_{\varepsilon\varepsilon}$ along this line in Fourier space. We will use this as a starting point for a perturbation theory, allowing us to determine $G_{\varepsilon\varepsilon}$ close to this line in Fourier space. From this we will find that in the low $T$ limit (with $k^2/T$ fixed), the Green's function near this line typically exhibits a single pole with dispersion relation \eqref{eq:scalingdispersion}. Exceptional cases arise near particular points on this line, for which a more sophisticated perturbation theory will be presented in the following section.

First we choose a point $(\omega_*,k_*)$ in Fourier space that lies on the line $\Omega=0$, around which we will perturbatively evaluate $G_{\varepsilon\varepsilon}$. Note that although we will often write $\omega_*$ and $k_*$ independently in the equations that follow, they are not really independent but are constrained by $\Omega(\omega_*,k_*)=0$. It will be simplest to think of us choosing a wavenumber $k_*$ which then fixes $\omega_*(k_*)$.

We now take frequencies and wavenumbers that are close to this point in Fourier space
\begin{equation}
\label{eq:perturbativefreqansatz}
\omega=\omega_*+\delta\omega,\quad\quad\quad\quad k^2=k_*^2+\delta (k^2),
\end{equation}
and look perturbatively for a solution to the equation \eqref{eq:axionEFeqn} of the form
\begin{equation}
\label{eq:perturbativepsiansatz}
\psi(r,\omega,k)=\psi(r,\omega_*,k_*)+\delta\psi(r,\omega_*,k_*),
\end{equation}
where we will evaluate $\delta\psi(r,\omega_*,k_*)$ to first order in $\delta\omega$ and $\delta(k^2)$. It will sometimes be convenient for us to repackage the two parameters $\delta\omega$ and $\delta(k^2)$ as $\delta\omega$ and $\delta\Omega$ where
\begin{equation}
\label{eq:deltaOmegadefn}
\delta\Omega=(2r_0^2-m^2)3ir_0\delta\omega+(2k_*^2+m^2)\delta (k^2),
\end{equation}
follows from \eqref{eq:Omegadefn}.

Substituting \eqref{eq:perturbativefreqansatz} and \eqref{eq:perturbativepsiansatz} into the equation \eqref{eq:axionEFeqn} yields
\begin{equation}
\label{eq:LOeqn}
\frac{d}{dr}\left[\frac{r^2fe^{-2i\omega_*r_*}}{(k_*^2+r^3f')^2}\partial_r\psi(r,\omega_*,k_*)\right]=0,
\end{equation}
at leading order. The first correction obeys
\begin{equation}
\begin{aligned}
\label{eq:NLOeqn}
\frac{d}{dr}\left[\frac{r^2fe^{-2i\omega_*r_*}}{(k_*^2+r^3f')^2}\partial_r\delta\psi(r,\omega_*,k_*)\right]=&\;2\frac{r^2fe^{-2i\omega_*r_*}\partial_r\psi(r,\omega_*,k_*)}{\left(k_*^2+r^3f'\right)^2}\frac{d}{dr}\left[i\delta\omega r_*+\frac{\delta (k^2)}{k_*^2+r^3f'}\right]\\
&+\delta\Omega\frac{e^{-2i\omega_*r_*}}{r^2\left(k_*^2+r^3f'\right)^3}\psi(r,\omega_*,k_*).
\end{aligned}
\end{equation}
The remaining task is to solve these equations with appropriate boundary conditions.

\subsubsection{Leading order solution}

The leading order equation \eqref{eq:LOeqn} can be integrated trivially for any $k_*$, giving
\begin{equation}
\label{eq:LOgeneralsol}
\psi(r,\omega_*,k_*)=C_0+C_1\int dr\frac{e^{2i\omega_*r_*}\left(k_*^2+r^3f'\right)^2}{r^2f},
\end{equation}
where $C_0$ and $C_1$ are constants. Imposing the ingoing boundary conditions means that we require that $\psi(r,\omega_*,k_*)$ has a Taylor series expansion near the horizon.

Near the horizon, the exponential term in the integrand $e^{2i\omega_*r_*}\sim (r-r_0)^{\frac{i\omega_*}{2\pi T}}$ (which for generic $\omega_*$ corresponds to an outgoing mode). Thus for generic $\omega_*$, the ingoing solution is
\begin{equation}
\label{eq:PsiLOGeneric}
\psi(r,\omega_*,k_*)=1.
\end{equation}
Substituting this into the expression \eqref{eq:axionG} we obtain the leading order Green's function
\begin{equation}
G_{\varepsilon\varepsilon}=\frac{k_*^2(k_*^2+m^2)}{i\omega_*}.
\end{equation}
This leading order result is exact for points lying exactly on the line $\Omega=0$ and the fact that it is finite mean that there are no poles lying exactly on this line when $\omega_*\ne0$.

Although the ingoing solution for generic choices of $k_*$ is given by \eqref{eq:PsiLOGeneric}, there are important exceptional cases. Assuming that $k_*^2+r_0^3f'(r_0)\ne0$, the integrand of \eqref{eq:LOgeneralsol} is of the form $\sim (r-r_0)^{\frac{i\omega_*}{2\pi T}-1}$ near the horizon. Therefore for the exceptional choices of wavenumber $k_*^2=k_n^2$ such that
\begin{equation}
\label{eq:exceptionalaxionpoints}
\omega_*(k_n)=\omega_n=-i2\pi T(n+2),\quad\quad n=-1,0,1,2,\ldots,
\end{equation}
the integral term in \eqref{eq:LOgeneralsol} will also obey the appropriate ingoing boundary condition. For these cases, the leading order ingoing solution is not simply that in \eqref{eq:PsiLOGeneric}, and we will address them separately in the next section. There are two additional exceptional cases that we will not discuss as they have been studied extensively elsewhere: $k_*^2=-r_0^3f'(r_0)\implies \omega_*=+i2\pi T$ \cite{Blake:2018leo}, and the usual hydrodynamic expansion $k_*^2=\omega_*=0$ \cite{Davison:2014lua}.

\subsubsection{First order correction when $\text{Im}(\omega)>-2\pi T$}

Returning now to the generic case, we substitute the leading order ingoing solution \eqref{eq:LOgeneralsol} into \eqref{eq:NLOeqn} to obtain the following equation for the first correction
\begin{equation}
\frac{d}{dr}\left[\frac{r^2fe^{-2i\omega_*r_*}}{(k_*^2+r^3f')^2}\partial_r\delta\psi(r,\omega_*,k_*)\right]=\delta\Omega\frac{e^{-2i\omega_*r_*}}{r^2\left(k_*^2+r^3f'\right)^3}.
\end{equation}
This can be trivially integrated to give the general solution
\begin{equation}
\label{eq:deltapsigeneralsol}
\delta\psi(r,\omega_*,k_*)=\delta\Omega\int dr\left(\frac{(k_*^2+r^3f')^2e^{2i\omega_*r_*}}{r^2f}\int dr\frac{e^{-2i\omega_*r_*}}{r^2\left(k_*^2+r^3f'\right)^3}\right).
\end{equation}
Each integral in this expression gives rise to an integration constant that we must fix to maintain the appropriate ingoing boundary conditions.

Consider first the inner integrand. Recalling that near the horizon $e^{-2i\omega_*r_*}\sim (r-r_0)^{-\frac{i\omega_*}{2\pi T}}$, the inner integral is only finite at the horizon for $\text{Im}(\omega_*)>-2\pi T$. We will assume this condition for now, and will show in the next section that it can be relaxed without affecting the results. With this condition, we write the inner integral as
\begin{equation}
\label{eq:innerintegralrewrite}
\int dr\frac{e^{-2i\omega_*r_*}}{r^2\left(k_*^2+r^3f'\right)^3}=C_2+\int_{r_0}^r dr\frac{e^{-2i\omega_*r_*}}{r^2\left(k_*^2+r^3f'\right)^3},
\end{equation}
where $C_2$ is a constant. Substituting this into \eqref{eq:deltapsigeneralsol}, we deduce that $C_2=0$ in order for $\delta\psi$ to obey ingoing boundary conditions at the horizon. Therefore the ingoing solution is given by
\begin{equation}
\delta\psi(r,\omega_*,k_*)=\delta\Omega\int_{r_0}^r dr\left(\frac{(k_*^2+r^3f')^2e^{2i\omega_*r_*}}{r^2f}\int_{r_0}^r dr\frac{e^{-2i\omega_*r_*}}{r^2\left(k_*^2+r^3f'\right)^3}\right),
\end{equation}
where we have also imposed the second boundary condition $\psi(r_0,\omega,k)=1$.

We can now evaluate the energy density Green's function using equation \eqref{eq:axionG}. From our point of view the important piece is the term $(-r^2\psi'+i\omega\psi)_{r\rightarrow\infty}$ in the denominator, which contains the information about poles of $G_{\varepsilon\varepsilon}$. At $\omega=\omega_*+\delta\omega$ and $k=k_*+\delta k$, this term has the form
\begin{equation}
\begin{aligned}
\label{eq:genericGresult}
(-r^2\psi'+i\omega\psi)_{r\rightarrow\infty}=&\;i\omega_*+i\delta\omega-\delta\Omega(k_*^2+m^2)^2\int_{r_0}^\infty dr\frac{e^{-2i\omega_*r_*}}{r^2\left(k_*^2+r^3f'\right)^3}\\
&\;+i\omega_*\delta\Omega\int_{r_0}^\infty dr\left(\frac{(k_*^2+r^3f')^2e^{2i\omega_*r_*}}{r^2f}\int_{r_0}^r dr\frac{e^{-2i\omega_*r_*}}{r^2\left(k_*^2+r^3f'\right)^3}\right).
\end{aligned}
\end{equation}

While the perturbative corrections in \eqref{eq:genericGresult} are formally small, the key point is that in an appropriate low temperature limit one of these corrections becomes anomalously large and thus is important. This allows us to deduce that at low $T$ the Green's function near the $\Omega=0$ line in Fourier space is dominated by a single pole whose dispersion relation we will shortly compute.

The relevant low $T$ limit is $T\rightarrow0$, keeping $k_*^2/T$ fixed (and therefore $\omega_*/T$ fixed), for which the integral on the first line of \eqref{eq:genericGresult} diverges like $1/T^2$. The divergence arises from the near-horizon region of the integral, and to isolate it we define the AdS$_2$ radial coordinate $\zeta$ as in equation \eqref{eq:axionAdS2coords}, rescale $T\rightarrow\epsilon\,T$, $k_*^2\rightarrow\epsilon\,k_*^2$, $\omega_*\rightarrow\epsilon\,\omega$, and then expand the integrand in the $\epsilon\rightarrow0$ limit to obtain
\begin{equation}
\begin{aligned}
\label{eq:IRintegralextraction}
\int_{r_0}^r dr\frac{e^{-2i\omega_*r_*}}{r^2\left(k_*^2+r^3f'\right)^3}&\;=\;\frac{1}{\epsilon^2}\int^\zeta_0d\zeta\frac{4\sqrt{2}3^{\frac{5}{2}-\frac{i\omega_*}{2\pi T}}\zeta^{-\frac{i\omega_*}{2\pi T}}\left(4\pi T+3\zeta\right)^{\frac{i\omega_*}{2\pi T}}}{m^2\left(\sqrt{6}k_*^2+m4\pi T+6m\zeta\right)^3}+\ldots\\
&\;=\;\frac{1}{\epsilon^2}\frac{12\sqrt{2}\left(4\pi T+3\zeta\right)^23^{\frac{3}{2}-\frac{i\omega_*}{2\pi T}}\left(\frac{\zeta}{4\pi T+3\zeta}\right)^{1-\frac{i\omega_*}{2\pi T}}}{m^3\left(16\pi^2T^2+4\omega_*^2\right)\left(\sqrt{6}k_*^2+m4\pi T+6m\zeta\right)^2}+\ldots.
\end{aligned}
\end{equation}
Evaluating this integral up to a UV cutoff $\Lambda$ of the near-horizon region (with $\Lambda\gg T$), we find the following divergent, cutoff-independent contribution to the integral on the first line of \eqref{eq:genericGresult}
\begin{equation}
\label{eq:axdivergentIRresult}
\int_{r_0}^\infty dr\frac{e^{-2i\omega_*r_*}}{r^2\left(k_*^2+r^3f'\right)^3}=\frac{1}{\epsilon^{2}} \frac{3\sqrt{6}}{4m^5\left(2\pi T+i\omega_*\right)(2\pi T-i\omega_*)}+\ldots.
\end{equation}
The ellipsis denotes terms that are subleading in the $\epsilon\rightarrow0$ limit. The double integral in equation \eqref{eq:genericGresult} also diverges like $1/\epsilon^2$ in the same limit.

In the expression \eqref{eq:genericGresult} for the denominator of $G_{\varepsilon\varepsilon}$, we observe the first and third terms will dominate in the low $T$ limit described above. More specifically, if we scale $\delta\Omega\rightarrow\epsilon^3\delta\Omega$ and $\delta\omega\rightarrow\epsilon^3\delta\omega$ along with the scalings of $T,\omega_*$ and $k_*^2$ above, then in the limit $\epsilon\rightarrow0$ the denominator is
\begin{equation}
\label{InvGepsGeps2}
\begin{aligned}
(-r^2\psi'+i\omega\psi)_{r\rightarrow\infty}=&\;\epsilon\left(i\omega_*-\delta\Omega\frac{3\sqrt{6}}{4m\left(2\pi T+i\omega_*\right)(2\pi T-i\omega_*)}\right)+\ldots.
\end{aligned}
\end{equation}
The right hand side vanishes for a suitable $\delta\Omega$, which means that $G_{\varepsilon\varepsilon}$ is dominated by a pole at the corresponding location in the complex $\omega$ plane. More specifically, for a given fixed wavenumber $k=k_*$ (i.e.~$\delta(k^2)=0$) we can replace $\delta\Omega$ by $\delta\omega$ using equation \eqref{eq:deltaOmegadefn} and then solve for $\delta\omega(\omega_*)$. Recalling that $\omega_*$ and $k_*$ are related by $\Omega(\omega_*,k_*)=0$, we obtain the pole location
\begin{equation}
\label{eq:expandeddispersionresult}
\omega(k)=-i\epsilon\sqrt{\frac{3}{2}}\frac{k^2}{m}\left(1+\epsilon\frac{k^2}{m^2}+\epsilon^2\left(\frac{4\pi T^2}{3m^2}+\frac{k^4}{m^4}\right)+\ldots\right).
\end{equation}
quoted in equation \eqref{eq:scalingdispersion} in the main text. 

The result \eqref{eq:expandeddispersionresult} contains information about the low $T$ limit of the hydrodynamic dispersion relation.  For example, by comparing \eqref{eq:expandeddispersionresult} to \eqref{eq:axionexactD} we would obtain the correct low temperature expansion of the diffusivity. But the result \eqref{eq:expandeddispersionresult} is not simply the hydrodynamic expansion of the dispersion relation: the terms with higher powers of $k^2$ are not being ignored but are formally subleading in the expansion we have described. Furthermore, although we have assumed that $\text{Im}(\omega)>-2\pi T$ until now, we will shortly see that the result \eqref{eq:expandeddispersionresult} continues to hold beyond this. In fact, even for values of $k$ outside the radius of convergence of the hydrodynamic expansion of the dispersion relation, we will see that there is still a pole at the location indicated by equation \eqref{eq:expandeddispersionresult}.

\subsubsection{Extension to general $\text{Im}(\omega)$}

In our derivation of the result \eqref{eq:expandeddispersionresult} we assumed that $\text{Im}(\omega)>-2\pi T$ in order that we could rewrite an integral appearing in the perturbative solution for $\delta\psi$ as \eqref{eq:innerintegralrewrite}. When this inequality is not obeyed, we must take more care as the integral formally diverges near the horizon. To deal with this, we will manually separate out the divergent part of the integral before imposing ingoing boundary conditions in an analogous manner to the steps above. The upshot of this is that $G_{\varepsilon\varepsilon}$ continues to be dominated by a pole with dispersion relation \eqref{eq:expandeddispersionresult} independently of the value of $\text{Im}(\omega)$ (within the low $T$ limit under consideration). The only exceptions to this are near the points \eqref{eq:exceptionalaxionpoints} that we previously highlighted, which we address in the following section.

To extend the results to lower in the complex $\omega$ plane, we write the inner integral appearing in the solution \eqref{eq:deltapsigeneralsol} for $\delta\psi$ as
\begin{equation}
\int\frac{dre^{-2i\omega_*r_*}}{r^2(k_*^2+r^3f')^3}=\int dr\left(r-r_0\right)^{-i\omega_*/2\pi T}G(r),
\end{equation}
where
\begin{equation}
G(r)=\frac{\left(r^2+rr_0+r_0^2-\frac{m^2}{2}\right)^{i\omega_*/4\pi T}}{r^2(k_*^2+r^3f')^3}\left(\frac{2r+r_0+\sqrt{2m^2-3r_0^2}}{2r-r_0-\sqrt{2m^2-3r_0^2}}\right)^{-\frac{i\omega_*(m^2-3r_0^2)}{4\pi Tr_0\sqrt{2m^2-3r_0^2}}}.
\end{equation}
We can formally expand $G(r)$ in a Taylor series near the horizon with coefficients $G_n$ and then write the integral as
\begin{equation}
\begin{aligned}
\label{eq:extensioneq1}
\int\frac{dre^{-2i\omega_*r_*}}{r^2(k_*^2+r^3f')^3}=&\;\int dr\Biggl\{\frac{e^{-2i\omega_*r_*}}{r^2(k_*^2+r^3f')^3}-\sum_{n=0}^NG_n\left(r-r_0\right)^{n-\frac{i\omega_*}{2\pi T}}\Biggr\}\\
&\;+\sum_{n=0}^N\frac{G_n}{1+n-\frac{i\omega_*}{2\pi T}}\left(r-r_0\right)^{1+n-\frac{i\omega_*}{2\pi T}},
\end{aligned}
\end{equation}
where the integral on the right hand side is finite provided that $\text{Im}\left(\omega_*\right) >-2\pi T\left(2+N\right)$. 

With this expression we now evaluate the contribution of the $\delta\psi'$ term in the perturbative expansion of the Greens function denominator \eqref{eq:axionG}, anticipating that it will again be anomalously large in the relevant low $T$ limit, and find
\begin{equation}
\begin{aligned}
\lim_{r\rightarrow\infty}\left(-r^2\delta\psi'\right)=-\left(m^2+k_*^2\right)^2\delta\Omega\lim_{r\rightarrow\infty}\Biggl[&\sum_{n=0}^N\frac{G_n}{(1+n-\frac{i\omega_*}{2\pi T})}(r-r_0)^{1+n-\frac{i\omega_*}{2\pi T}}\\
&+\int_{r_0}^rdr\left\{\frac{e^{-2i\omega_*r_*}}{r^2(k_*^2+r^3f')^3}-\sum_{n=0}^NG_n(r-r_0)^{n-\frac{i\omega_*}{2\pi T}}\right\}\Biggr].
\end{aligned}
\end{equation}
The terms on the first line will vanish provided that $\text{Im}\left(\omega_*\right)<-2\pi T(1+N)$. Assuming this, along with the previous condition, we find a contribution to the Green's function denominator given by
\begin{equation}
\lim_{r\rightarrow\infty}\left(-r^2\delta\psi'\right)=-\left(m^2+k_*^2\right)^2\delta\Omega\int_{r_0}^\infty dr\left\{\frac{e^{-2i\omega_*r_*}}{r^2(k_*^2+r^3f')^3}-\sum_{n=0}^NG_n(r-r_0)^{n-\frac{i\omega_*}{2\pi T}}\right\},
\end{equation}
for $-2\pi T(2+N)<\text{Im}(\omega_*)<-2\pi T(1+N)$. There are an extra set of terms in comparison to the corresponding term in the Green's function denominator \eqref{eq:genericGresult} for $\text{Im}(\omega_*)>-2\pi T$.

By setting $N=0,1,2,\ldots$ and evaluating this integral, we can extend our previous results to successively lower regions of the complex $\omega$ plane. To isolate the divergent contribution to the integral arising from the near-horizon region we follow the same procedure as before: take the low $T$, near-horizon scaling limit of the integrand and integrate the leading term up to a UV cutoff $\Lambda\gg T$ of the near-horizon region. After doing this explicitly for $N=0,1,2,3$ we find the same result \eqref{eq:axdivergentIRresult} as before, and we expect that this will continue to be the case for higher $N$. As a consequence, the results for the Green's function derived previously (including the dispersion relation \eqref{eq:expandeddispersionresult} for the pole) are also valid in the region $\text{Im}(\omega)<-2\pi T$.

\subsection{Analytic description of pole collisions}

We now turn to the exceptional points \eqref{eq:exceptionalaxionpoints} for which the analysis in the previous section must be modified. Recall that for these points the leading order ingoing solution is not given by \eqref{eq:PsiLOGeneric}, despite the fact that they lie on the line $\Omega=0$. In fact, at these points the ingoing solution is no longer uniquely defined and thus the perturbative expansion is more subtle. These `pole-skipping points' correspond to points where a line of poles of the field theory Green's function intersects with a line of zeroes \cite{Grozdanov:2017ajz,Blake:2017ris,Blake:2018leo,Blake:2019otz}. Here, we will take a direct approach and just describe in practice how to obtain the appropriate perturbative solutions.

The main result of this calculation is that in the low $T$ limit, there are in fact two poles of $G_{\varepsilon\varepsilon}$ in the vicinity of each exceptional point with $n\ge0$. One passes directly through the point while the other is separated from it by a distance $\sim T^2$ in $\omega$ space. These two poles collide for specific complex values of $k$, that become real as $T\rightarrow0$. Assuming that the collision occurring at the smallest value of $\left|k\right|$ characterizes the breakdown of diffusive hydrodynamics (which we will confirm numerically in the next section), this allows us to quantitatively extract $\omega_{eq}$ and $k_{eq}$ (or equivalently $\tau_{eq}$ and $v_{eq}$) in the low $T$ limit.

\subsubsection{Leading order ingoing solution}

Our starting point is the result \eqref{eq:LOgeneralsol} for the general solution for $\psi$, at a point $(\omega_*,k_*)$ lying exactly on the line $\Omega(\omega_*,k_*)=0$. At the frequencies $\omega_n$ given in \eqref{eq:exceptionalaxionpoints}, this solution obeys ingoing boundary conditions at the horizon for any values of the constants. The corresponding values of $k_n$ are found by solving $\Omega(\omega_n,k_n)=0$. This equation has two distinct solutions for $k_n^2$ and ultimately we will be interested in the one with $k_n^2\sim T$ at low $T$. Using equation \eqref{eq:axionT} to trade $T$ and $r_0$, explicitly this is
\begin{equation}
\begin{aligned}
\label{eq:kndefn}
k_n^2&\,=\frac{1}{2}\left(-m^2+\sqrt{-m^4(5+3n)+24m^2(2+n)r_0^2-36(2+n)r_0^4}\right)\\
&\,= \sqrt{\frac{8}{3}}\left(2+n\right)m\pi T-\frac{8}{3}\left(2+n\right)^2\pi^2T^2+O(T^3),
\end{aligned}
\end{equation}
where on the second line we have written the small $T$ expansion.

To fix the boundary conditions at these points we must consider points in Fourier space infinitesimally away from $(\omega_n,k_n)$, where the ingoing solution is unique. Specifically, we take the equation of motion \eqref{eq:axionEFeqn} at frequencies $\omega=\omega_n+\delta\omega$ and wavenumbers $k^2=k_n^2+\delta(k^2)$ and construct a Taylor series solution for $\psi$ near the horizon. This solution is ingoing by definition, and is unique (up to an overall normalisation). We subsequently expand the Taylor series solution at small $\delta\omega\sim\delta (k^2)$, finding the leading order result
\begin{equation}
\begin{aligned}
\label{eq:PoleSkippingNH1}
\psi(r,\omega_n+\delta\omega,k_n+\delta k)\rightarrow1+c_n\frac{\delta\Omega}{\delta\omega}(r-r_0)^{2+n}+\ldots,
\end{aligned}
\end{equation}
where the coefficients $c_n$ are independent of both $\delta\omega$ and $\delta(k^2)$. The important aspect of this solution is that, although it is unique, it depends on the ratio $\delta\Omega/\delta\omega$ that parameterizes exactly how we have moved away from the point $(\omega_n,k_n)$ in Fourier space. By expanding our general leading order solution \eqref{eq:LOgeneralsol} near the horizon and matching it to the ingoing solution \eqref{eq:PoleSkippingNH1}, we obtain the values of the constants $C_0$ and $C_1$ that correspond to an ingoing solution at the point $(\omega_n+\delta\omega,k_n+\delta k_n)$. Explicitly, the ingoing solution at leading order is then
\begin{equation}
\label{eq:PoleSkippingNH2}
\psi(r,\omega_n,k_n)=1+i2\pi T\alpha_{n}^{(1)}\frac{\delta\Omega}{\delta\omega}I_{n}^{(1)}(r),
\end{equation}
where we have defined the constants
\begin{equation}
\alpha_n^{(j)}=\frac{1}{(2+n-j)!}\left.\left(\frac{d^{2+n-j}}{dr^{2+n-j}}\left(\frac{(r-r_0)^{2+n}e^{-4\pi T(2+n)r_*}}{r^2\left(k_{n}^2+r^3f'\right)^3}\right)\right)\right|_{r=r_0},
\end{equation}
and the integrals
\begin{equation}
I_n^{(j)}(r)\equiv\int^r_{r_0}dr\frac{e^{2i\omega_n r_*}\left(k_{n}^2+r^3f'\right)^2}{r^2f(r-r_0)^{j-1}}.
\end{equation}

\subsubsection{Correction to the leading order solution}

Having established the leading order solution \eqref{eq:PoleSkippingNH2} near these special points in Fourier space, we will now repeat the same procedure at one higher order in the expansion at small $\delta\omega\sim\delta(k^2)$ to obtain the first correction to this result. After substituting the leading order result \eqref{eq:PoleSkippingNH2} into the equation \eqref{eq:NLOeqn}, we can trivially integrate the resulting equation to obtain the expression
\begin{equation}
\begin{aligned}
\label{eq:intereq1}
\delta\psi(r,\omega_n,k_n)=&\;\int dr\left(\frac{e^{2i\omega_nr_*}(k_n^2+r^3f')^2}{r^2f}\delta\Omega\int dr\left(-\frac{4\pi T\alpha_n^{(1)}}{r^2f}+\frac{e^{-2i\omega_nr_*}}{r^2\left(k_n^2+r^3f'\right)^3}\right)\right)\\
&\;+i2\pi T\alpha_n^{(1)}\frac{\delta\Omega^2}{\delta\omega}\int^r_{r_0}dr\left(\frac{e^{2i\omega_nr_*}(k_n^2+r^3f')^2}{r^2f}\int_{r_0}^rdr\frac{e^{-2i\omega_nr_*}I_n^{(1)}(r)}{r^2\left(k_n^2+r^3f'\right)^3}\right)\\
&\;+i4\pi T\alpha_n^{(1)}\frac{\delta\Omega\delta (k^2)}{\delta\omega}\int^r_{r_0}dr\frac{e^{2i\omega_nr_*}(k_n^2+r^3f')}{r^2f},
\end{aligned}
\end{equation}
for the correction. There are two arbitrary integration constants arising from the two indefinite integrals in the first term, and these need to be fixed in accordance with the ingoing boundary conditions. 

Before doing this, note that for all $n\ge0$ the inner integral of the first term diverges near the horizon and so it will be convenient to separate out the divergent terms by hand to leave a finite integral.\footnote{\baselineskip1pt For the special case $n=-1$, the summation terms in \eqref{eq:sumsumsum} and subsequent equations can be set to zero.} Specifically, we re-write it as
\begin{equation}
\begin{aligned}
\label{eq:sumsumsum}
\int dr\Biggl(\frac{-4\pi T\alpha_n^{(1)}}{r^2f}+\frac{e^{-2i\omega_nr_*}}{r^2\left(k_n^2+r^3f'\right)^3}\Biggr)=&\;\int^r_{r_0}dr\Biggl(\frac{e^{-2i\omega_nr_*}}{r^2\left(k_n^2+r^3f'\right)^3}-\frac{4\pi T\alpha_n^{(1)}}{r^2f}-\sum_{j=2}^{2+n}\frac{\alpha_{n}^{(j)}}{(r-r_0)^j}\Biggr)\\
&\;+\beta_n+\sum_{j=2}^{2+n}\frac{\alpha_{n}^{(j)}}{(1-j)(r-r_0)^{j-1}},
\end{aligned}
\end{equation}
where $\beta_n$ is the integration constant. Substituting this into \eqref{eq:intereq1}, we obtain the solution in the form
\begin{equation}
\begin{aligned}
\label{eq:deltapsin1}
\delta\psi(r,\omega_n,k_n)=&\;\delta\Omega\Biggl\{\int^r_{r_0}dr\Biggl(\frac{e^{2i\omega_nr_*}(k_n^2+r^3f')^2}{r^2f}\int^r_{r_0}dr\left(\frac{e^{-2i\omega_nr_*}}{r^2\left(k_n^2+r^3f'\right)^3}-\frac{4\pi T\alpha_n^{(1)}}{r^2f}-\sum_{j=2}^{2+n}\frac{\alpha_{n}^{(j)}}{(r-r_0)^j}\right)\\
&\;+\beta_n^{(0)}I_n^{(1)}(r)+\sum_{j=2}^{2+n}\frac{\alpha_n^{(j)}}{(1-j)}I_n^{(j)}(r)\Biggr)\Biggr\}\\
&+\frac{\delta\Omega^2}{\delta\omega}\Biggl\{i2\pi T\alpha_n^{(1)}\int^r_{r_0}dr\left(\frac{e^{2i\omega_nr_*}(k_n^2+r^3f')^2}{r^2f}\int_{r_0}^rdr\frac{e^{-2i\omega_nr_*}I_n^{(1)}(r)}{r^2\left(k_n^2+r^3f'\right)^3}\right)+\beta_n^{(1)}I_n^{(1)}(r)\Biggr\}\\
&+\frac{\delta\Omega\delta (k^2)}{\delta\omega}\Biggl\{i4\pi T\alpha_n^{(1)}\int^r_{r_0}dr\frac{e^{2i\omega_nr_*}(k_n^2+r^3f')}{r^2f}+\beta_n^{(2)}I_n^{(1)}(r)\Biggr\}.
\end{aligned}
\end{equation}
In this expression we fixed one integration constant such that $\psi(r_0,\omega,k)=1$. The remaining integration constant $\beta_n$ has been split up into parts proportional to $\delta\Omega,\delta\Omega^2/\delta\omega$ and $\delta\Omega\delta (k^2)/\delta\omega$ for later convenience.

The integration constants $\beta_n^{(i)}$ are fixed by imposing ingoing boundary conditions. To determine these conditions explicitly we again use the unique Taylor series solution for $\psi(r,\omega_n+\delta\omega,k_n+\delta k)$ near the horizon, and find the correction to equation \eqref{eq:PoleSkippingNH1} at the next order in the small $\delta\omega\sim\delta (k^2)$ expansion. At this order, the series has the form
\begin{equation}
\label{eq:deltapsiNHingoing}
\psi(r,\omega_n+\delta\omega,k_n+\delta k)=0+\ldots+(r-r_0)^{2+n}\left[\gamma_n^{(0)}\delta\Omega+\gamma_n^{(1)}\frac{\delta\Omega^2}{\delta\omega}+\gamma_n^{(2)}\frac{\delta\Omega\delta (k^2)}{\delta\omega}\right]+\ldots.
\end{equation}
The coefficients $\gamma_n^{(i)}$ become increasingly complicated for higher values of $n$ and so we will only explicitly present the $n=0$ expressions, which are those relevant for the breakdown of hydrodynamics
\begin{equation}
\begin{aligned}
\gamma_0^{(0)}=\:&\frac{4m^2}{r_0^2(6r_0^2-m^2)^2(2k_n^2+6r_0^2-m^2)}-\frac{12(m^2-2r_0^2)}{r_0^2(6r_0^2-m^2)(2k_n^2+6r_0^2-m^2)^2},\\
\gamma_0^{(1)}=\:&-\frac{2i}{r_0^3(6r_0^2-m^2)(2k_n^2+6r_0^2-m^2)^2},\\
\gamma_0^{(2)}=\:&\frac{18i(m^2-2r_0^2)}{r_0^3(2k_n^2+6r_0^2-m^2)^3}-\frac{4i(m^2-3r_0^2)}{r_0^3(6r_0^2-m^2)(2k_n^2+6r_0^2-m^2)^2}.
\end{aligned}
\end{equation}
To impose ingoing boundary conditions on the solution \eqref{eq:deltapsin1}, we expand it near the horizon and match it to the ingoing solution \eqref{eq:deltapsiNHingoing} to fix the values of the three integration constants $\beta_n^{(i)}$. Explicitly, these integration constants are then
\begin{equation}
\begin{aligned}
\label{eq:betadefns}
\beta_n^{(0)}=\frac{n+2}{p_n}\gamma_n^{(0)}-\sum_{j=2}^{2+n}\frac{\sigma_n^{(j-1)}\alpha_{n}^{(j)}}{p_n(1-j)},\quad\quad\beta_n^{(1)}=\frac{n+2}{p_n}\gamma_n^{(1)},\quad\quad
\beta_n^{(2)}=\frac{n+2}{p_n}\gamma_n^{(2)}-\frac{i4\pi T\alpha_n^{(1)}}{k_n^2+r_0^3f'(r_0)},
\end{aligned}
\end{equation}
where
\begin{equation}
\begin{aligned}
p_n=\frac{\left(k_n^2+r_0^3f'(r_0)\right)^2}{(4\pi T)^{2+n/2}r_0^{1+n/2}}\left(\frac{3r_0+\sqrt{2m^2-3r_0^2}}{3r_0-\sqrt{2m^2-3r_0^2}}\right)^{\frac{(2+n)(m^2-3r_0^2)}{2r_0\sqrt{2m^2-3r_0^2}}},
\end{aligned}
\end{equation}
and
\begin{equation}
\begin{aligned}
\sigma_n^{(j)}=\frac{1}{j!}\left.\left(\frac{d^{j}}{dr^{j}}\left(\frac{e^{4\pi T(2+n)r_*}(k_n^2+r^3f')^2}{r^2f(r-r_0)^{1+n}}\right)\right)\right|_{r=r_0}.
\end{aligned}
\end{equation}
In summary, the first correction to the ingoing solution in the small $\delta\omega\sim\delta (k^2)$ expansion is given by \eqref{eq:deltapsin1} and \eqref{eq:betadefns}.

\subsubsection{Low temperature limit of the Green's function}

The ingoing solution near $(\omega_n,k_n)$, including the first correction in the small $\delta\omega\sim\delta (k^2)$ expansion, is given by the sum of \eqref{eq:PoleSkippingNH1} and \eqref{eq:deltapsin1}. It is convenient to rewrite this as
\begin{equation}
\psi(r,\omega_n+\delta\omega,k_n+\delta k)=1+\Phi_n^{(0)}(r)\frac{\delta\Omega}{\delta\omega}+\Phi_n^{(1)}(r)\delta\Omega+\Phi_n^{(2)}(r)\frac{\delta\Omega^2}{\delta\omega}+\ldots,
\end{equation}
where the ellipsis represents higher order terms in the small $\delta\omega\sim\delta\Omega$ expansion, and
\begin{equation}
\begin{aligned}
\Phi_n^{(0)}(r)=&\;i2\pi T\alpha_n^{(1)}I_n^{(1)}(r),\\
\Phi_n^{(1)}(r)=&\;\left(\beta_n^{(0)}-\frac{3ir_0(2r_0^2-m^2)}{2k_n^2+m^2}\beta_n^{(2)}\right)I_n^{(1)}(r)+\sum_{j=2}^{2+n}\frac{\alpha_{n}^{(j)}}{(1-j)}I_n^{(j)}(r)\\
&\;\;\;+\int^r_{r_0}dr\left(\frac{e^{2i\omega_nr_*}(k_n^2+r^3f')^2}{r^2f}\int^r_{r_0}dr\left(\frac{e^{-2i\omega_nr_*}}{r^2\left(k_n^2+r^3f'\right)^3}-\frac{4\pi T\alpha_n^{(1)}}{r^2f}-\sum_{j=2}^{2+n}\frac{\alpha_{n}^{(j)}}{(r-r_0)^j}\right)\right)\\
&+\frac{12\pi Tr_0(2r_0^2-m^2)\alpha_n^{(1)}}{2k_n^2+m^2}\int^r_{r_0}dr\frac{e^{2i\omega_nr_*}(k_n^2+r^3f')}{r^2f},\\
\Phi_n^{(2)}(r)=&\;\left(\beta_n^{(1)}+\frac{\beta_n^{(2)}}{2k_n^2+m^2}\right)I_n^{(1)}(r)+\frac{i4\pi T\alpha_n^{(1)}}{2k_n^2+m^2}\int^r_{r_0}dr\frac{e^{2i\omega_nr_*}(k_n^2+r^3f')}{r^2f}\\
&+i2\pi T\alpha_n^{(1)}\int^r_{r_0}dr\left(\frac{e^{2i\omega_nr_*}(k_n^2+r^3f')^2}{r^2f}\int_{r_0}^rdr\frac{e^{-2i\omega_nr_*}I_n^{(1)}(r)}{r^2\left(k_n^2+r^3f'\right)^3}\right).
\end{aligned}
\end{equation}
Substituting into \eqref{eq:axionG}, the denominator of the energy density Green's function near $(\omega_n,k_n)$ is
\begin{equation}
\begin{aligned}
\label{eq:GnearPSpt}
&(-r^2\psi'+i\omega\psi)_{r\rightarrow\infty}\;\propto\lim_{r\rightarrow\infty}\Biggl[\delta\Omega\left(-r^2{\Phi_n^{(0)}}'(r)+i\omega_n\Phi_n^{(0)}(r)\right)+2\pi T(2+n)\delta\omega+i\delta\omega^2\\
&+\delta\Omega^2\left(-r^2{\Phi_n^{(2)}}'(r)+i\omega_n\Phi_n^{(2)}(r)\right)+\delta\omega\delta\Omega\left(-r^2{\Phi_n^{(1)}}'(r)+i\Phi_n^{(0)}(r)+i\omega_n\Phi_n^{(1)}(r)\right)\Biggr].
\end{aligned}
\end{equation}

The first important result can be seen from this formal expression: it vanishes as $\delta\omega,\delta\Omega\rightarrow0$ and therefore there is always a pole of the Green's function passing through the point $(\omega_n,k_n)$. The dispersion relation of the pole in the immediate vicinity of this point can be found by taking the $O(\delta\omega)$ and $O(\delta\Omega)$ terms in \eqref{eq:GnearPSpt} and solving for $\delta\omega$ as a function of $\delta (k^2)$.

As the $O(\delta^2)$ terms are formally small, one might naively conclude that the Green's function is dominated by this single pole in the vicinity of $(\omega_n,k_n)$. However the second important result is that, in the low $T$ limit, the coefficient of the $\delta\Omega\delta\omega$ term becomes anomalously large for all $n\ge0$. This indicates that there is an additional pole of the Green's function near $(\omega_n,k_n)$ in these cases. By carefully evaluating the expression \eqref{eq:GnearPSpt} for the Green's function at low $T$ we will further be able to show that for each $n\ge0$ these two poles undergo a collision in complex Fourier space near $(\omega_n,k_n)$. The collision closest to the origin of $k$ space ($n=0$) involves the hydrodynamic diffusion mode and thus characterizes the breakdown of hydrodynamics.

To take the low temperature limit, we formally rescale $T\sim\epsilon$ and then evaluate the scaling of each term in \eqref{eq:GnearPSpt} in the limit $\epsilon\rightarrow0$.\footnote{\baselineskip1pt To accurately extract divergent near-horizon contributions in this limit we used the procedure described around equation \eqref{eq:IRintegralextraction}.} For the $n=-1$ case, the coefficient of the $\delta\Omega$ term in \eqref{eq:GnearPSpt} is $\sim T^{-1}$ in this limit while the coefficients of the $\delta\omega\delta\Omega$ and $\delta\Omega^2$ terms are both $\sim T^{-2}$. This is what one might have naively expected: the quadratic corrections are small provided $\delta\omega,\delta\Omega\lesssim T$ and thus the Green's function is dominated by a single pole over a region of Fourier space of size $\omega\sim T$ near $\omega=-i2\pi T$. This is consistent with what we found for a generic point near $\Omega=0$.

However, for the cases $n\geq0$, the scaling of the coefficients in the low $T$ limit is different: the coefficient of the $\delta\Omega$ term is $\sim T^0$ while the coefficients of the $\delta\omega\delta\Omega$ and $\delta\Omega^2$ terms are $\sim T^{-2}$ and $\sim T^{-1}$ respectively. Therefore the quadratic corrections are important at frequencies $\delta\omega\sim T^2$, indicating that the single-pole approximation to the Green's function breaks down here due to the presence of a second pole. In the limit $T\rightarrow0$, these two poles are parametrically closer to one another in Fourier space than the typical separation $\sim T$ between poles.

To make this more quantitative, we can formally rescale $\delta\Omega\sim\delta\omega\sim\epsilon^2$ and then compute \eqref{eq:GnearPSpt} in the low $\epsilon\rightarrow0$ limit for $n\geq0$. If we keep only the leading order terms in this limit ($O(\epsilon^2)$), the Green's function denominator near $(\omega_n,k_n)$ has the simple factorised form
\begin{equation}
\begin{aligned}
\label{eq:LOcollisionGR}
 \delta\Omega\left(1-i\tau_n\delta\omega\right)\propto \left(\mathcal{D}_n\delta(k^2)-i\delta\omega\right)\left(1-i\tau_n\delta\omega\right),
\end{aligned}
\end{equation}
where we replaced $\delta\Omega$ using \eqref{eq:deltaOmegadefn}, $\mathcal{D}_n=\sqrt{3/2}m^{-1}$, and $\tau_n$ is given in equation \eqref{eq:taundefn} in the main text.\footnote{\baselineskip1pt We computed $\mathcal{D}_n,\tau_n,\lambda_n$ explicitly for $n=0,1,2,3$ and then extrapolated to general $n$.} The expression \eqref{eq:LOcollisionGR} clearly shows that the Green's function near the points $(\omega_n,k_n)$ with $n\geq0$ is dominated by two poles. Within this approximation the poles will coincide at the exactly imaginary frequency $\delta\omega=-i\tau_n^{-1}$ and real wavenumber $\delta(k^2)=\tau_n^{-1}\mathcal{D}_n^{-1}$. But this is not yet really a pole collision\footnote{\baselineskip1pt More precisely, the dispersion relation does not have a branch point here.}: the poles pass through each other undisturbed, and the dispersion relations $\delta\omega(\delta (k^2))$ extracted from \eqref{eq:LOcollisionGR} have an infinite radius of convergence.

To observe the finite radius of convergence characteristic of a pole collision, we must go beyond the result \eqref{eq:LOcollisionGR} and include subleading corrections in the $\epsilon$ expansion. One source of these are finite $T$ corrections to $\tau_n$ and $\mathcal{D}_n$. Although these corrections are the largest in magnitude (e.g.~the leading correction to $\tau_n$ is suppressed by $T\log T$), we will ignore them as they will not alter the functional form of \eqref{eq:LOcollisionGR} and so will not produce a pole collision. The other source of corrections are the $\delta\omega$, $\delta\Omega^2$ and $\delta\omega^2$ terms present in the full expression \eqref{eq:GnearPSpt}. The first two corrections enter at $O(\epsilon^3)$ while the latter is $O(\epsilon^4)$. Of the first two, we will only need to consider that proportional to $\delta\omega$ as it is clear from inspecting the first line of \eqref{eq:LOcollisionGR} that adding a term $\propto\delta\Omega^2$ will still result in two nicely factorised poles whose dispersion relations $\delta\omega(\delta (k^2))$ each have an infinite radius of convergence. Upon adding the leading $\delta\omega$ term correction to \eqref{eq:LOcollisionGR}, we obtain the expression
\begin{equation}
\label{eq:appendixGnearcollision}
G_{\varepsilon\varepsilon}^{-1}\left(\omega,k\right)\propto\left(\mathcal{D}_n\delta(k^2)-i\delta\omega\right)\left(1-i\tau_n\delta\omega\right)-i\lambda_n\delta\omega,
\end{equation}
for the energy density Green's function denominator presented in the main text. 

The expression \eqref{eq:appendixGnearcollision} no longer factorizes in a simple manner and solving explicitly for the dispersion relations $\delta\omega(\delta(k^2))$ of the two poles results in expressions with a non-zero discriminant term. The poles will collide at the wavenumber where the discriminant vanishes, which is when
\begin{equation}
\delta (k_c^2)=\mathcal{D}_n^{-1}\tau_n^{-1}\left(1\pm i\lambda_n^{1/2}\right)^2\implies\delta\omega_c=-i\tau_n^{-1}\left(1\pm i\lambda_n^{1/2}\right).
\end{equation}
Expanding out these expressions for $\delta\omega_c$ and $\delta (k_c^2)$ at low $T$, and recalling that $\omega=\omega_n+\delta\omega$ and $k^2=k_n^2+\delta(k^2)$, gives the collision location 
\begin{equation}
\begin{aligned}
\label{eq:fullcollisionlocations}
\omega_c=&\;-i2\pi T(2+n)\left(1+\frac{8\sqrt{6}\pi T}{9m}+\ldots\pm i\left(\frac{2^{13/4}}{3^{5/4}}\sqrt{(2+n)^2-1}\left(\frac{\pi T}{m}\right)^{3/2}+\ldots\right)\right)\\
k_c^2=&\; k_n^2\left(1+\frac{8\sqrt{6}\pi T}{9m}+\ldots\pm i\left(\frac{2^{17/4}}{3^{5/4}}\sqrt{(2+n)^2-1}\left(\frac{\pi T}{m}\right)^{3/2}+\ldots\right)\right),
\end{aligned}
\end{equation}
for every $n\geq0$. We have shown the leading order real and imaginary terms in each expression, and the ellipses denote small $T$ corrections to these. Note however, that there are corrections to the leading real (imaginary) term in $k_c^2$ ($\omega_c$) that are larger than the leading order imaginary (real) term (for example, coming from $T\log T$ suppressed corrections to $\tau_n$). 

There are two collision points in the complex $\omega$ plane with opposite signs of $\text{Re}(\omega)$ and these collision points asymptotically approach the imaginary (real) axis as $T\rightarrow0$. From \eqref{eq:fullcollisionlocations}, the phase of the collision wavenumber and frequency are
\begin{equation}
\begin{aligned}
\label{eq:axionphasecollisionappendix}
\phi_{k}=\frac{2^{13/4}\sqrt{(2+n)^2-1}}{3^{5/4}}\left(\frac{\pi T}{m}\right)^{3/2}+\ldots,\quad\quad\quad \phi_{\omega}=-\frac{\pi}{2}+\phi_{k},
\end{aligned}
\end{equation}
where we define these quantities using the collision in the upper right quadrant of the complex $k_c$ plane.

In summary, we have derived an analytic description of a series of pole collisions happening in the vicinity of the points $(\omega_n,k_n)$ for all $n\geq0$. The $n=0$ collision characterizes the breakdown of diffusive hydrodynamics (as can be seen from the numerical results in the following section) and from \eqref{eq:fullcollisionlocations} the local equilibration scales are therefore given by
\begin{equation}
\begin{aligned}
\label{eq:appendixaxioncollisionscales}
&\omega_{eq}=4\pi T\left(1+\frac{8\sqrt{6}\pi T}{9m}+\ldots\right),\quad\quad\quad k_{eq}^2=\frac{\omega_{eq}}{D_{\varepsilon}}\left(1-\frac{4\sqrt{6}\pi T}{3m}+\ldots\right),
\end{aligned}
\end{equation}
as quoted in equation \eqref{eq:axioncollisionscales} in the main text.

\subsection{Comparison with numerical results}

In this section we present exact results for the locations of poles of $G_{\varepsilon\varepsilon}$ obtained numerically. In addition to verifying our analytic results in the appropriate regions of Fourier space, these numerical results provide a global picture of the pole structure and therefore show how the different analytic results we have presented connect together.

To obtain the exact locations of poles of $G_{\varepsilon\varepsilon}$
one needs to solve for the relevant linear fluctuations around the black hole
geometry \eqref{eq:adsaxionappendix}. As explained above one can 
determine $G_{\varepsilon\varepsilon}$ by solving for a suitably defined variable
$\tilde \psi$ (see  \eqref{eq:linaxmastervble}) that satisfies a decoupled equation of motion.
For our numerical calculations we followed~\cite{Blake:2018leo}  and solved
instead for 
\begin{equation}
\Psi_1(r,\omega,k)={r^2\,f\over (k^2+r^3\,f')^2}\,{d\over dr}\left[
(k^2+r^3\,f')\,\tilde\psi(r,\omega,k)\right]\,,
\label{eq:linaxpsi1vble}
\end{equation}
which satisfies the equation of motion
\begin{equation}
{d\over dr}\left[{r^2\,f\,(k^2+r^3\,f')^3\over\omega^2(k^2+r^3\,f')
	-k^2(k^2+m^2\,f)}\Psi_1'\right]
+{(k^2+r^3\,f')^2\over r^2\,f}\,\Psi_1=0\,.
\label{eq:linaxpsi1eom}
\end{equation}
We solved this equation numerically using a dimensionless radial variable
$z=r_0/r$. We integrated from the horizon
at $z=1$, where we imposed ingoing boundary conditions
$\Psi_1= \psi_{h,0}\,(1-z)^{-i\omega/(4\pi\,T)}\left[1+O\left((1-z)\right)\right]$,
to the boundary at $z=0$ where we read off the leading and subleading
contributions of $\Psi_1= \psi_0+\psi_1\,z+O(z^2)$. We used 8 terms in the near-horizon expansion, an IR cutoff of $1-z=10^{-5}$ and working precision 50 in Mathematica.\footnote{\baselineskip1pt At the lowest temperatures and largest wavenumber we
decreased the IR  cutoff to $1-z=10^{-6}$ and increased the working precision to 60.} Finally, the poles of $G_{\varepsilon\varepsilon}$ were identified as zeroes
of the ratio $\psi_0/\psi_1$~\cite{Blake:2018leo}.

\begin{figure}
\begin{center}
\includegraphics[scale=0.75]{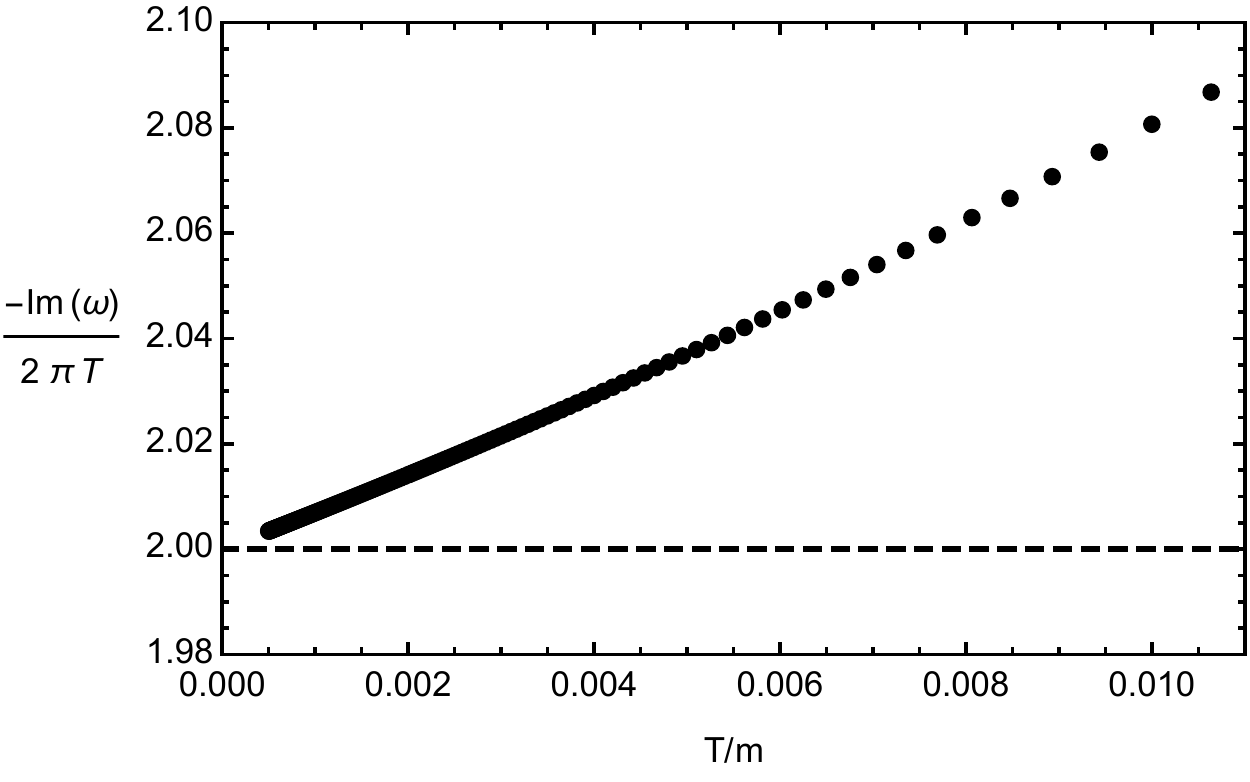}
\end{center}
\caption{Numerical results for the frequency of the longest lived non-hydrodynamic modes of the neutral, translation-symmetry breaking model at $k=10^{-1}T$. The dashed black line indicates the infra-red frequency $\omega_0$ defined in equation \eqref{eq:IRpolelocations}.}
\label{fig:axionIRmodelocations}
\end{figure}

For small wavenumbers $k$, it is the hydrodynamic pole of $G_{\varepsilon\varepsilon}$ with dispersion relation \eqref{eq:hydrodiffusion} that lies closest to the origin of the complex $\omega$ plane. In order to characterize the breakdown of hydrodynamics we must first identify the non-hydrodynamic poles lying near to the origin with which this pole could collide. At small enough $T/m$ (and for small real $k$), a family of poles with imaginary frequencies are the non-hydrodynamic poles closest to the origin. In Figure \ref{fig:axionIRmodelocations} we show the locations of the longest-lived such pole in $\omega$ space at small $k$, demonstrating agreement with \eqref{eq:IRpolelocations} as $T/m\rightarrow0$.
$G_{\varepsilon\varepsilon}$ of course exhibits other non-hydrodynamic poles besides these, but at low $T/m$ these are far from the origin and so will not play any role in the rest of our discussion.

We will now examine more carefully how these infra-red poles move in the complex $\omega$ plane as real $k$ is varied, and how they fit together with the diffusive hydrodynamic mode. In the low $T$ limit, we have established that there is generically a pole with the dispersion relation \eqref{eq:expandeddispersionresult} but that near each of the points $(\omega_{n\ge0},k_{n\ge0})$ there is an additional pole. Given that the infra-red modes lie at $\omega_{n\ge0}$ as $k\rightarrow0$, the simplest explanation would be that the infra-red modes have a very weak $k$-dependence and are themselves the additional poles. This explanation is correct, as is shown in Figure \ref{fig:axionoverviewallmodes}.
Note that while this Figure shows multiple instances of poles coming very close to one another, the collisions themselves cannot be seen as they occur at the complex values of $k$ \eqref{eq:fullcollisionlocations}.

The final part of our numerical analysis is to examine more closely the dynamics of the poles near $(\omega_0,k_0)$ where we expect a collision between the hydrodynamic diffusion mode and the longest-lived infra-red mode to occur, characterising the breakdown of hydrodynamics. Based on our calculations in the previous section, we expect the collision to occur for the complex value of the wavenumber \eqref{eq:fullcollisionlocations}. In Figure \ref{fig:axioncollisionphase} we show pole trajectories in the complex $\omega$ plane as $\left|k\right|$ is varied at fixed $T$ and phase of $k$, illustrating that two poles collide for an appropriate $\left|k\right|$.

\begin{figure}
\begin{center}
\includegraphics[scale=0.6]{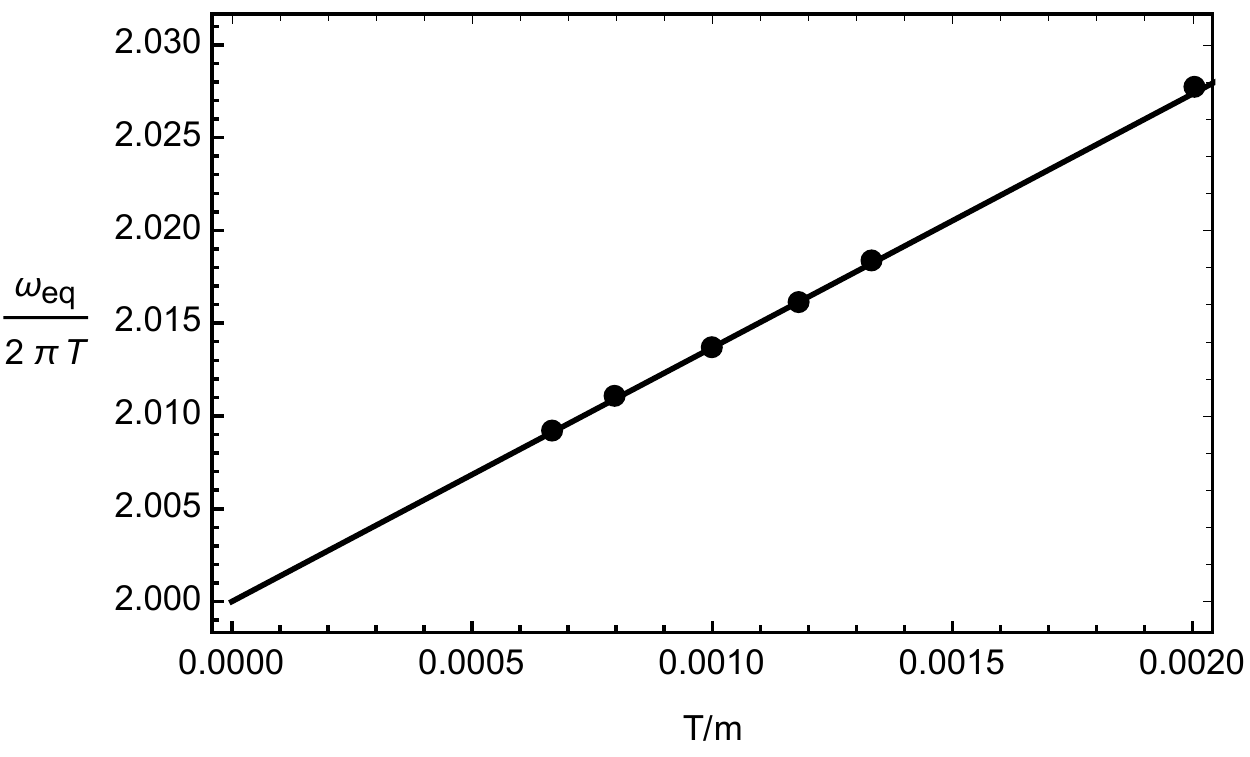}
\includegraphics[scale=0.58]{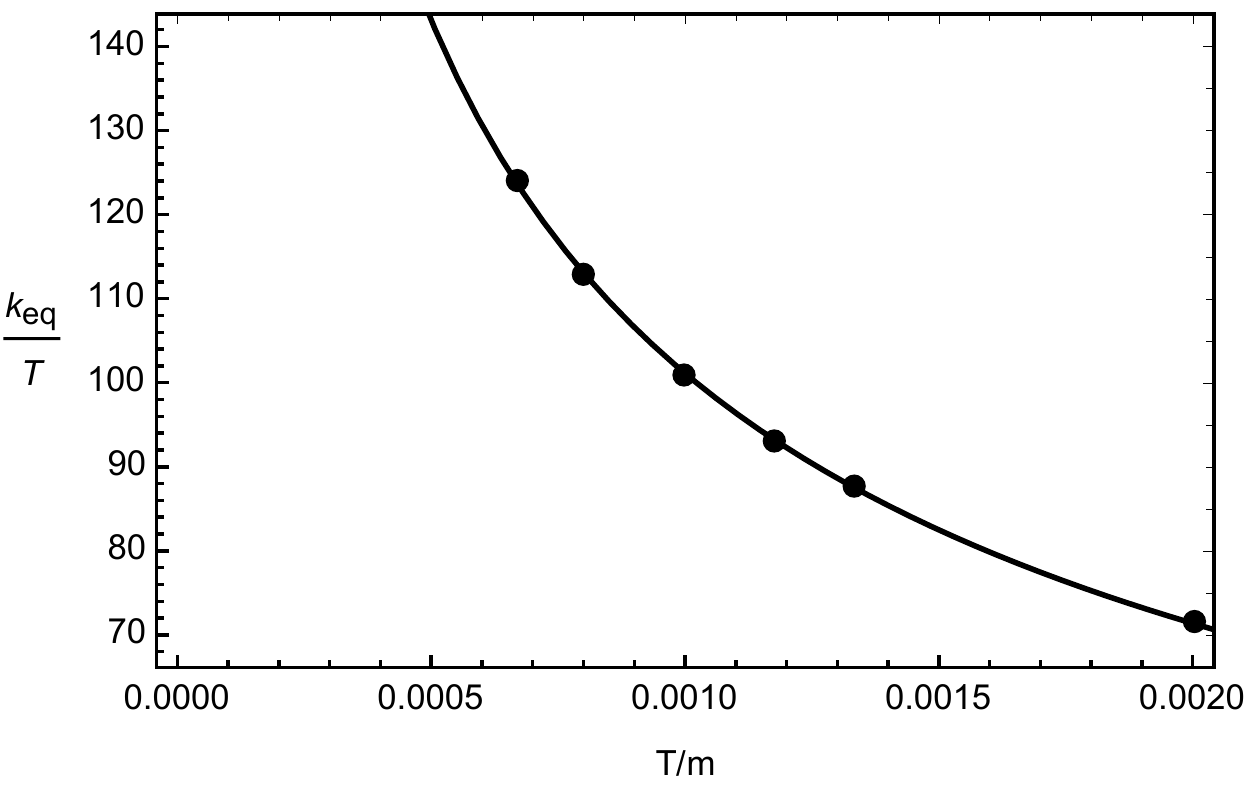}
\includegraphics[scale=0.6]{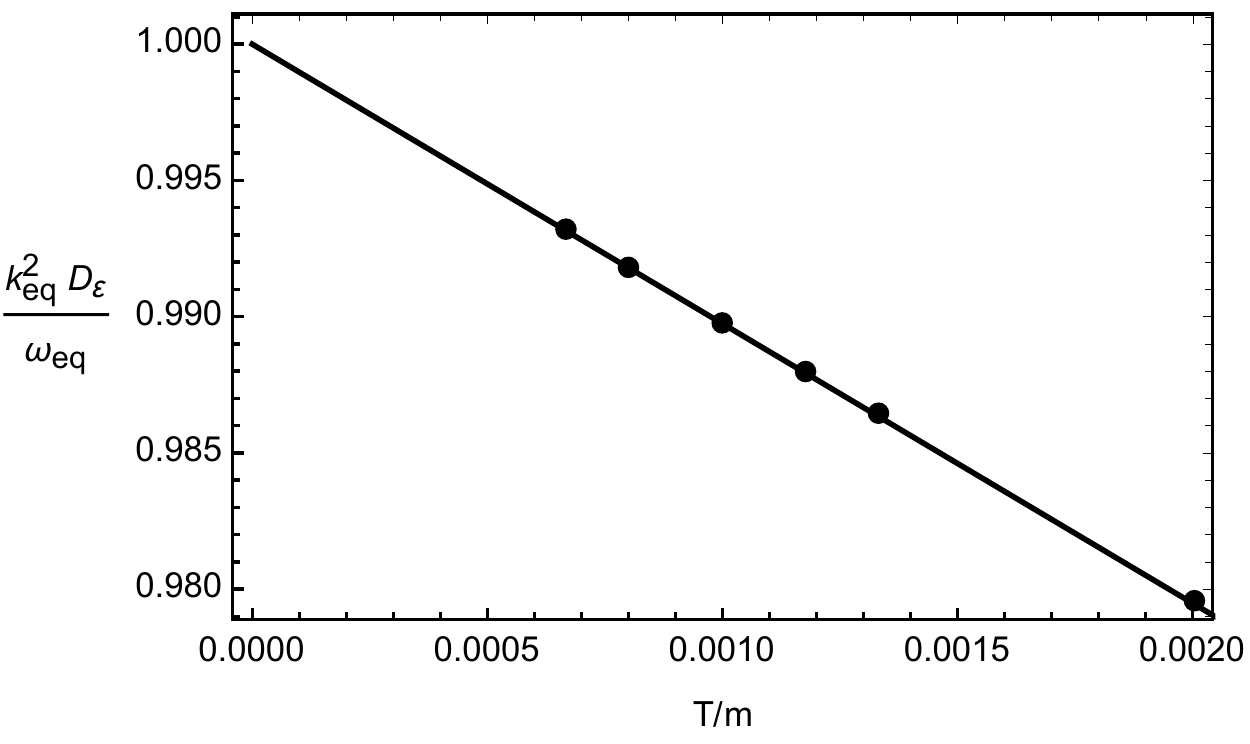}
\includegraphics[scale=0.59]{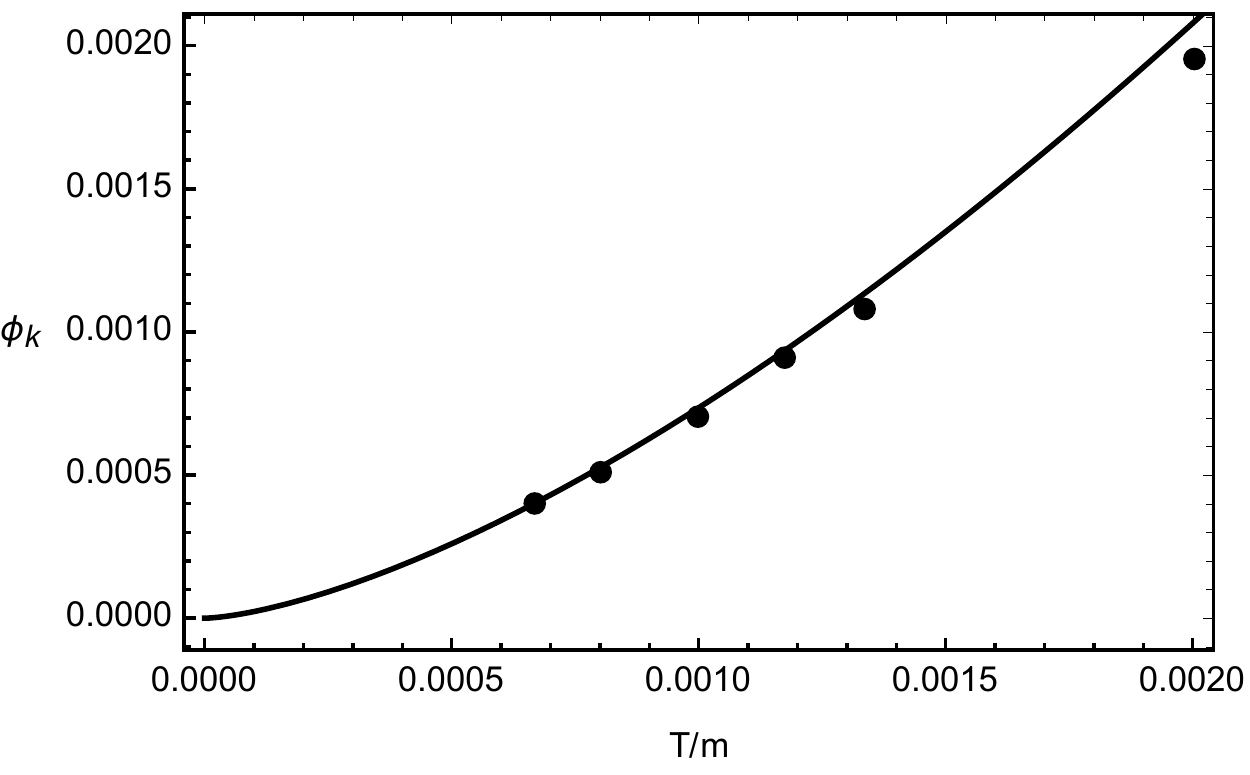}
\end{center}
\caption{Temperature dependence of the local equilibration data for the neutral, translational-symmetry breaking model obtained numerically (circles) and analytically in equations \eqref{eq:axionexactD}, \eqref{eq:axionphasecollisionappendix} and \eqref{eq:appendixaxioncollisionscales} (solid lines) from the $n=0$ pole collision.}
\label{fig:axionequilibrationdata}
\end{figure}

In Figure \ref{fig:axionequilibrationdata} we quantitatively compare features of this pole collision to the analytic results obtained in the previous section, showing excellent agreement as $T\rightarrow0$.
The upper two panels show exact results for the local equilibration scales $\omega_{eq}$ and $k_{eq}$, while the lower panels show the ratio $Dk_{eq}^2/\omega_{eq}$ and the phase of the collision wavenumber $\phi_k$.

In summary, the low $T$ numerical results show that as real $k$ is increased, the hydrodynamic diffusion mode moves down the imaginary $\omega$ axis before coming into the vicinity of the longest-lived infra-red mode near $\omega=-i4\pi T$. The breakdown of hydrodynamics is characterized by a collision between these two modes near this point when $k$ has a small imaginary part. The low $T$ collision is captured quantitatively by the analytic expressions derived in the previous section.

We finish this section with some observations on what happens after hydrodynamics breaks down. After the collision signaling the breakdown, one may expect the character of the collective modes of the system to change. But this does not happen. After the collision occurs, there continues to be a diffusive-type mode with dispersion relation \eqref{eq:expandeddispersionresult} as is apparent from Figure \ref{fig:axionoverviewallmodes} in the main text. In fact, this mode then undergoes a second collision near $\omega=-i6\pi T$ before quickly re-appearing and so on through a whole series of collisions near the frequencies $\omega_{n\ge0}$. So while hydrodynamics formally breaks down, diffusion-like transport continues to occur. 

\bigbreak

\section{Perturbations of AdS$_4$-Reissner-Nordstrom}
\label{sec:RNAppendix}

We study the AdS$_4$-RN spacetime as a solution to Einstein-Maxwell gravity with action
\begin{equation}
S=\int d^4x\sqrt{-g}\left(R+6-\frac{1}{4}F^2\right).
\end{equation}
The metric is given by
\begin{equation}
\begin{aligned}
\label{eq:adsRNappendix}
ds^2=-r^2f(r)dt^2+r^2\underline{dx}^2+\frac{dr^2}{r^2f(r)},\quad\quad\quad f(r)=1-\left(1+\frac{\mu^2}{4r_0^2}\right)\frac{r_0^3}{r^3}+\frac{\mu^2r_0^2}{4r^4},
\end{aligned}
\end{equation}
and is supported by the gauge field profile
\begin{equation}
A_t=\mu\left(1-\frac{r_0}{r}\right).
\end{equation}
The Hawking temperature $T$ is related to the horizon radius and chemical potential by
\begin{equation}
T=\frac{12r_0^2-\mu^2}{16\pi r_0}.
\end{equation}

\subsection{Gauge-invariant perturbations}

As for the neutral translation-breaking solution, the field theory Green's functions can be conveniently studied by defining four gauge-invariant combinations of the linearised perturbations that obey decoupled equations of motion \cite{Kodama:2003kk}. With appropriate conditions imposed at the AdS boundary, the quasinormal modes of two of these variables (`the longitudinal variables' $\psi^L_{\pm}$) correspond to the poles of the retarded Green's function of energy density $G_{\varepsilon\varepsilon}(\omega,k)$, while the quasinormal modes of the other two variables (`the transverse variables' $\psi^T_{\pm}$) correspond to poles of the retarded Green's function of transverse momentum density $G_{\Pi\Pi}(\omega,k)$ \cite{Edalati:2010hk,Edalati:2010pn}.

One can assign an infra-red dimension $\Delta(k)$ to each decoupled variable by examining its perturbation equation in the near-horizon AdS$_2$ region. For either pair of variables, the dimensions are \cite{Edalati:2010hk,Edalati:2010pn}
\begin{equation}
\Delta_{\pm}(k)=\frac{1}{2}\left(1+\sqrt{5+\frac{8k^2}{\mu^2}\pm4\sqrt{1+\frac{4k^2}{\mu^2}}}\right).
\end{equation}

This charged state supports three independent types of hydrodynamic mode: propagating sound waves, diffusion of energy and charge with diffusivity $D_\varepsilon$, and transverse diffusion of momentum with diffusivity $D_\Pi$. We focus on the latter two modes, whose diffusivities can be computed from Einstein relations \cite{Kovtun:2012rj} as
\begin{equation}
\begin{aligned}
\label{eq:RNdiffs}
D_{\varepsilon}=\frac{1}{3r_0}+\frac{8r_0}{3\left(\mu^2+4r_0^2\right)},\quad\quad\quad D_{\Pi}=\frac{4r_0}{3\left(\mu^2+4r_0^2\right)}.
\end{aligned}
\end{equation}

Each hydrodynamic mode corresponds to a quasinormal mode of one of the four decoupled bulk variables. The sound modes arise as quasinormal modes of the longitudinal variable $\psi^L_-$, diffusion of energy and charge arises as a quasinormal mode of the longitudinal variable $\psi^L_+$, and transverse diffusion of momentum arises as a quasinormal mode of the transverse variable $\psi^T_-$.\footnote{\baselineskip1pt Our definition of the variables $\psi^L_{\pm}$ is the opposite of that in \cite{Edalati:2010pn}, and instead agrees with that in \cite{Hartnoll:2012rj}.} From these observations, we obtain $\Delta_\varepsilon(0)=2$ and $\Delta_{\Pi}(0)=1$ for the $k\rightarrow0$ scaling dimensions of the variables that exhibit the respective hydrodynamic diffusion modes.

\subsection{Numerical results}

The linearised perturbation equations for the AdS$_4$-RN black brane solution are significantly more complicated than for the neutral translation-breaking model of Appendix \ref{sec:AxionAppendix}. We will therefore establish the results by solving these equations numerically to extract the relevant information about the poles of $G_{\varepsilon\varepsilon}$ and $G_{\Pi\Pi}$. The poles of the Green's functions have previously been studied in \cite{Edalati:2010hk,Edalati:2010pn,Brattan:2010pq,Ge:2010yc,Davison:2011uk,Davison:2013bxa,Moitra:2020dal} and more recently the breakdown of hydrodynamics was looked at in \cite{Withers:2018srf,Jansen:2020hfd}. The breakdown of hydrodynamics in the AdS$_5$ case was studied in \cite{Abbasi:2020ykq}.

To obtain our numerical results for the poles of $G_{\varepsilon\varepsilon}$ we used the coupled variables and equations described in \cite{Davison:2011uk}.\footnote{\baselineskip1pt Denoting the chemical potential in \cite{Davison:2011uk} by $\tilde{\mu}$, $\mu=2\tilde{\mu}$ due to a different normalisation of the action.} Although in these variables it is not manifest that the poles can be separated into two decoupled sectors, on occasion it will be useful for us to take advantage of this fact to simplify the results that we present. As for the perturbations of the neutral translation-breaking model we work with a dimensionless radial variable $z=r_0/r$  and integrated from the horizon at $z=1$ (where we imposed ingoing boundary conditions) to the boundary at $z=0$. To determine the poles of $G_{\varepsilon\varepsilon}$ we followed the determinant method of \cite{Kaminski:2009dh}. We used 12 terms in the near-horizon expansion, an IR cutoff of $1-z=10^{-5}$ and working precision $60$ in Mathematica. For the lowest temperatures and largest wavenumbers we worked with $1-z=10^{-6}$
and working precision $80$.

Our numerical results for the $G_{\Pi\Pi}$ were obtained by solving the perturbation equation for the decoupled variable $\psi^T_-$, as in \cite{Edalati:2010hk}. The poles arising from $\psi^T_+$ will not be relevant to the results we present. In this case, we used 9 terms in the near-horizon expansion, an IR cutoff of $1-z=10^{-5}$, and working precision 50 in Mathematica. As before, for the lowest temperatures and
largest wavenumbers we decreased the cutoff to $1-z=10^{-6}$ and increased the working precision to 70.

\begin{figure}
\begin{center}
\includegraphics[scale=0.80]{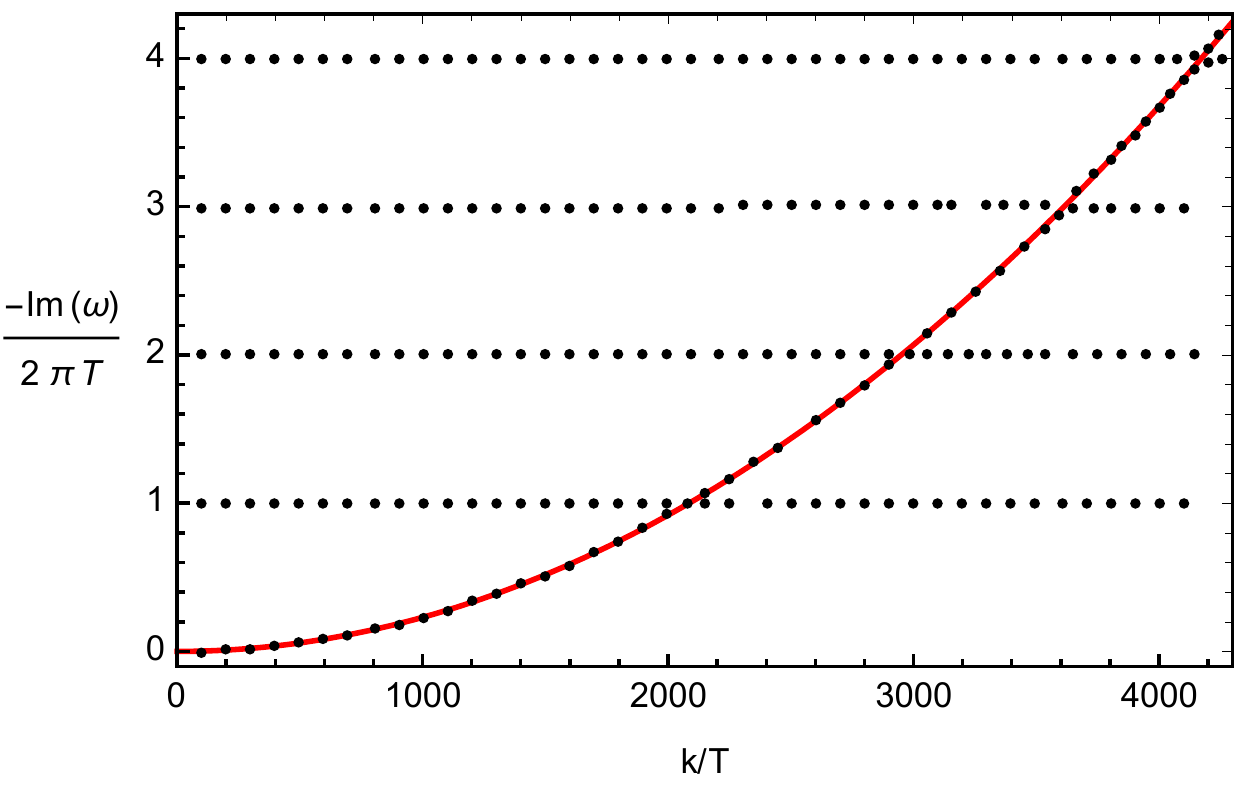}
\includegraphics[scale=0.80]{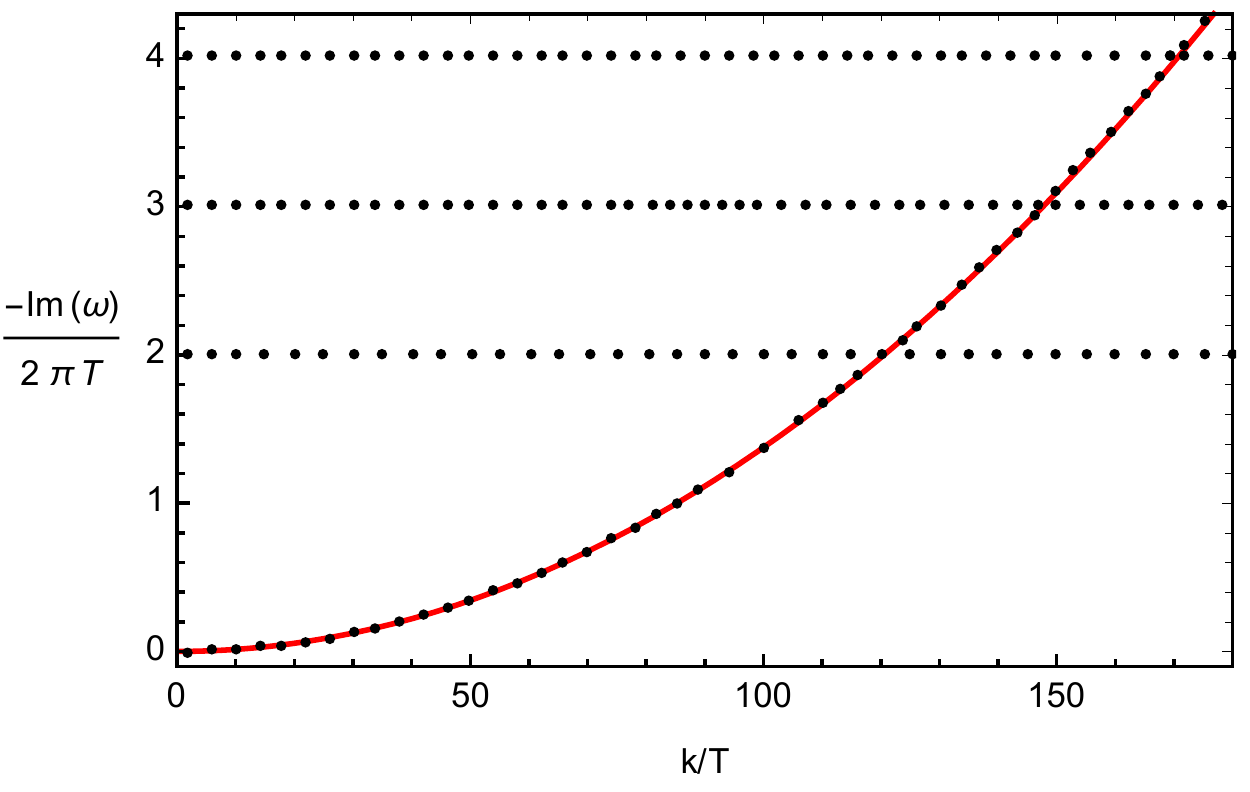}
\end{center}
\caption{Numerical results (black circles) for the locations of the relevant diffusive and infra-red modes of RN-AdS$_4$. The upper plot shows poles of $G_{\Pi\Pi}$ associated to the bulk variable $\psi^T_-$ at $T/\mu=5\times10^{-6}$ and the lower plot shows the poles of $G_{\varepsilon\varepsilon}$ associated to the bulk variable $\psi^L_+$ at $T/\mu=5\times10^{-4}$. Solid red lines show the quadratic approximation $\omega=-iD_{\varepsilon,\Pi}k^2$ to the hydrodynamic dispersion relation, with $D_{\varepsilon,\Pi}$ as in \eqref{eq:RNdiffs}.}
\label{fig:RNallmodes}
\end{figure}

For this state it is well-established that at low temperatures the longest-lived non-hydrodynamic poles of each Green's functions are a family of poles with imaginary frequencies $\sim T$ \cite{Edalati:2010hk,Edalati:2010pn}. In Figure \ref{fig:RNallmodes} we show the motion of the diffusive pole of each Green's function, along with that of the longest-lived non-hydrodynamic poles, as function of real $k$ for a fixed low temperature.
For both Green's functions the diffusive mode moves down the imaginary $\omega$ axis and is extremely well approximated by the quadratic approximation to diffusive hydrodynamics, while the infra-red modes are approximately $k$-independent and lie very close to the locations \eqref{eq:IRpolelocations} of the poles of the infra-red Green's functions. Note that $G_{\varepsilon\varepsilon}$ can be separated exactly into the sum of two independent pieces, each associated to one of the variables $\psi^L_\pm$. In Figure \ref{fig:RNallmodes}, we show only poles of $G_{\varepsilon\varepsilon}$ that are associated to the variable $\psi^L_+$. The reason is that generically a pole of $\psi^L_-$ cannot have any effect on the hydrodynamic diffusion pole (or its radius of convergence) even if these poles coincide, as they are solutions to independent differential equations.\footnote{\baselineskip1pt An exception to this arises at particular values of $k$ where the equations for the two decoupled variables become equivalent. Such collisions can be relevant to the breakdown of hydrodynamics \cite{Withers:2018srf,Abbasi:2020ykq,Jansen:2020hfd}, but not in the low $T$ regime which we study.}

\begin{figure}
\begin{center}
\includegraphics[scale=0.6]{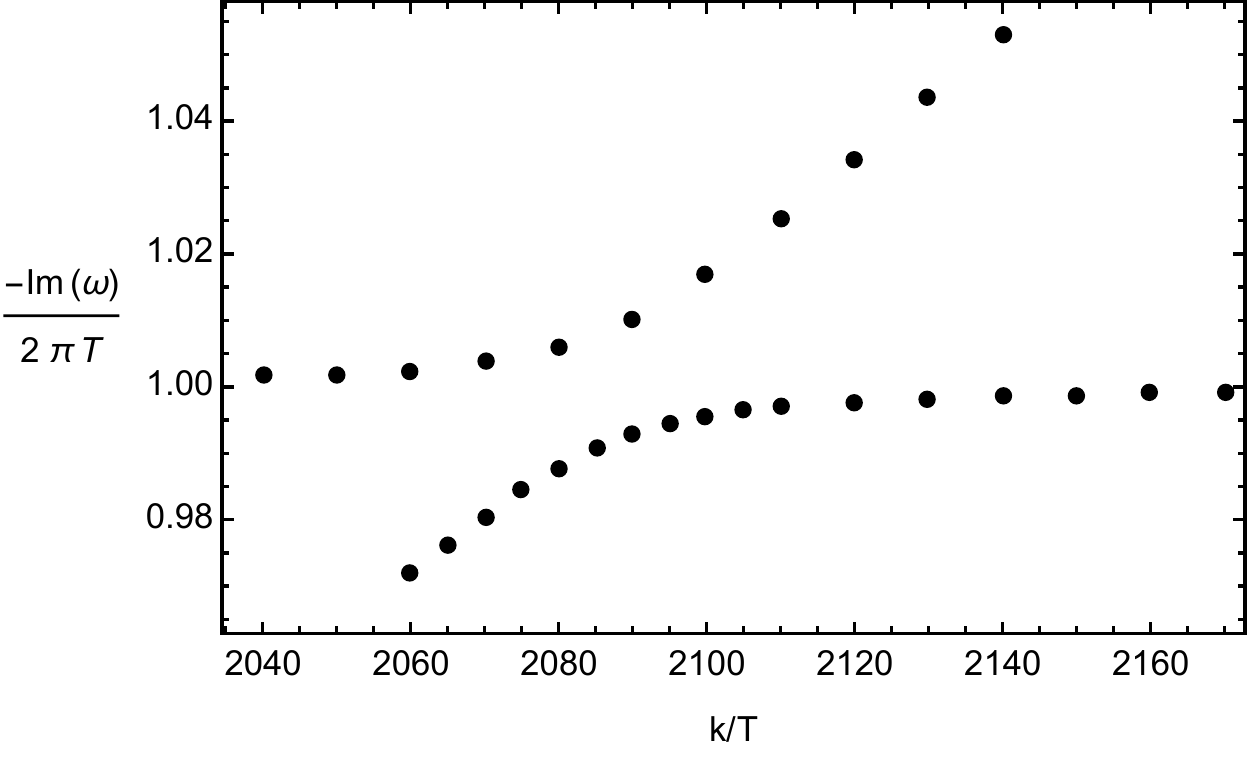}
\includegraphics[scale=0.6]{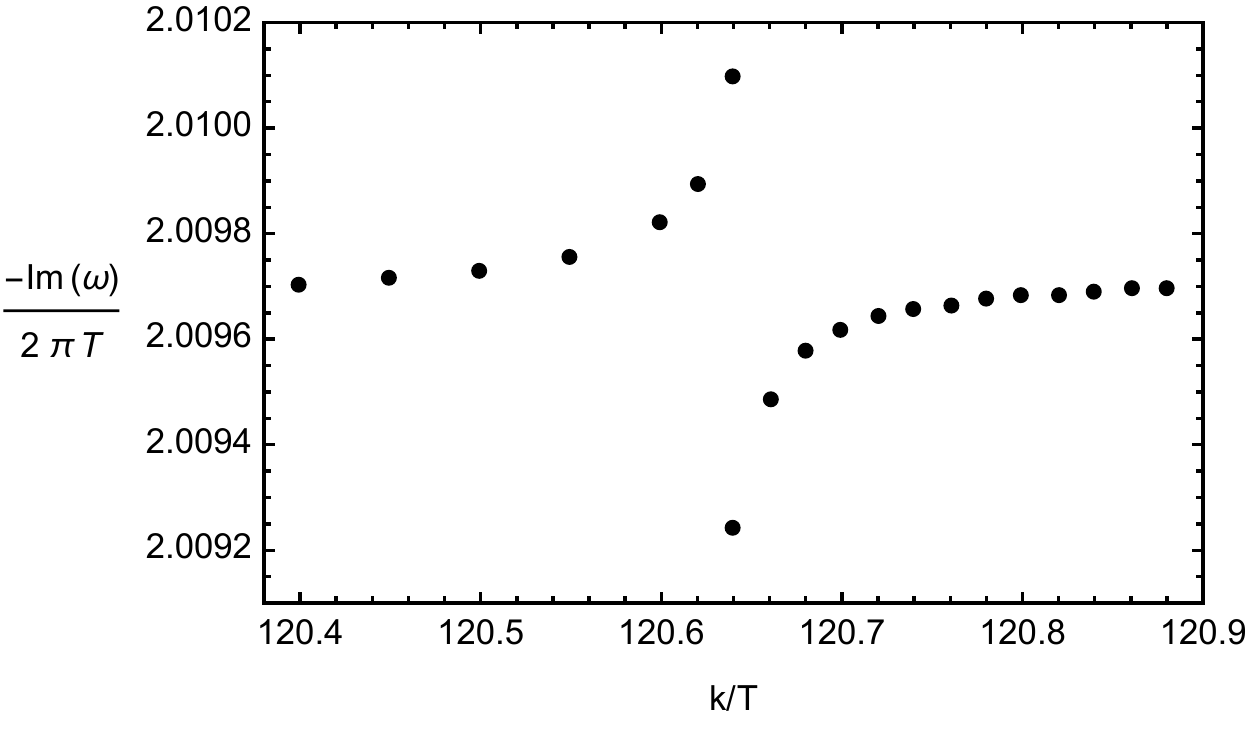}
\end{center}
\caption{Closeup of the numerical results for pole locations of $G_{\Pi\Pi}$ in the vicinity of $\omega=-i2\pi T$ at $T/\mu=5\times10^{-6}$ (left panel) and of $G_{\varepsilon\varepsilon}$ in the vicinity of $\omega=-i4\pi T$ at $T/\mu=5\times10^{-4}$ (right panel). There is no collision for real $k$ in either case.}
\label{fig:RNavoidedcrossing}
\end{figure}

In the vicinity of the frequencies $\omega_{n\ge0}$, the diffusive mode and an infra-red mode become very close, but a detailed inspection (Figure \ref{fig:RNavoidedcrossing}) shows that there are no collisions near these points for any real $k$.
To observe the pole collisions that characterize the breakdown of hydrodynamic diffusion, we must consider the motion of the poles for complex values of $k$. 

For $G_{\Pi\Pi}$, the first pole collision happens close to $\omega=-i2\pi T$ and plots of $\omega_{eq}$, $Dk_{eq}^2/\omega_{eq}$ and $\phi_k$ for this collision are shown in Figures \ref{fig:RNeqscalesmaintext} and \ref{fig:RNphasemaintext} in the main text. In \cite{Jansen:2020hfd} it was shown that $k_{eq}(T\rightarrow0)\rightarrow0$ (both by identifying the location of the pole collision we have described, and independently by explicitly computing the radius of convergence of the hydrodynamic series), which is consistent with the more quantitative results we have presented. At sufficiently high $T$, the nature of the collision characterising the breakdown of hydrodynamics undergoes a qualitative change \cite{Withers:2018srf,Jansen:2020hfd}.

\begin{figure}
\begin{center}
\includegraphics[scale=0.6]{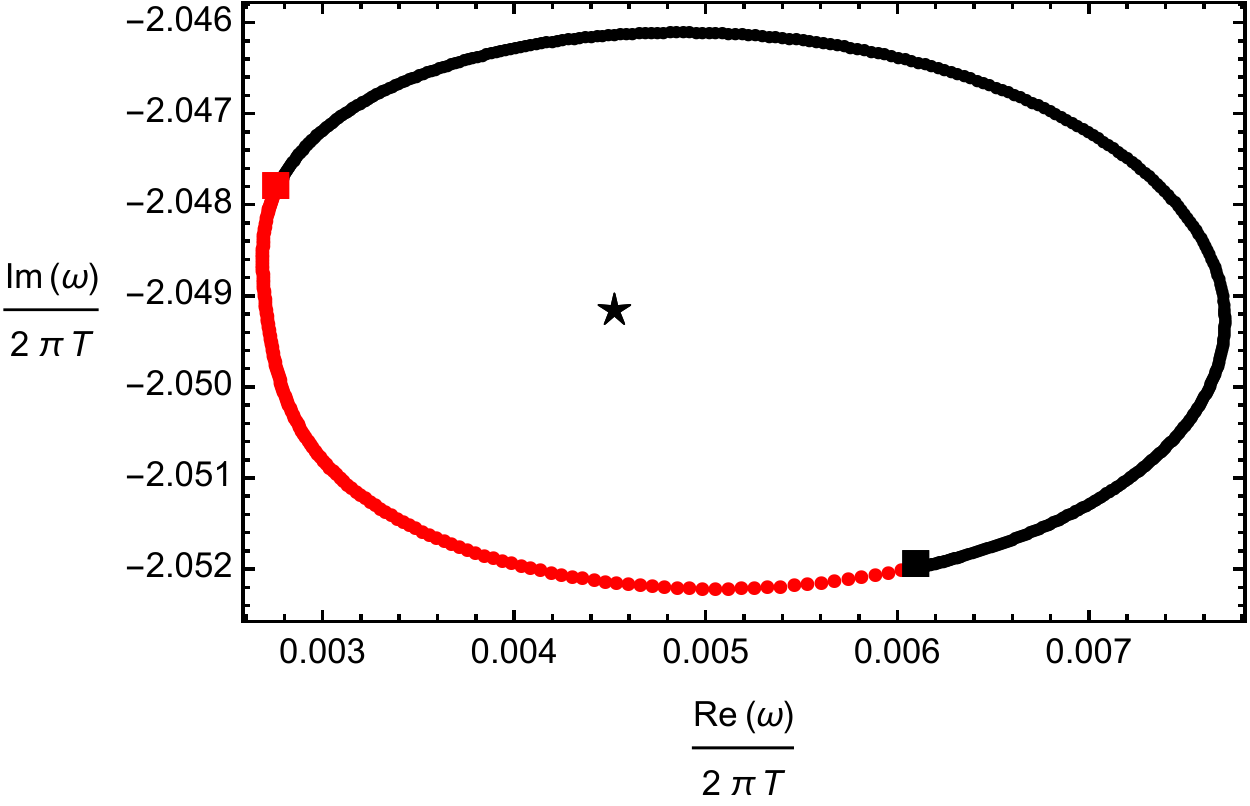}
\end{center}
\caption{Black and red dots show the locations of the two relevant modes of $G_{\varepsilon\varepsilon}$ for $T=0.005\mu$ as $k(\theta)/T\approx54.3+0.11i+0.02e^{i\theta}$, $0\leq\theta\leq2\pi$ moves around a small circle enclosing the collision wavenumber $k_c$. Squares show the location of each mode when $\theta=0$. The approximate location of the collision frequency $\omega_c$ is shown as a star. When $\theta=2\pi$ the two modes have exchanged places. They return to their original positions when $\theta=4\pi$.}
\label{fig:branchpoint}
\end{figure}

For $G_{\varepsilon\varepsilon}$ the breakdown of hydrodynamics is characterized by a collision near $\omega=-i4\pi T$ between the diffusive mode and the first infra-red mode of $\psi^L_+$. In Figures \ref{fig:RNeqscalesmaintext} and \ref{fig:RNphasemaintext} in the main text we show corresponding plots of $\omega_{eq}$, $Dk_{eq}^2/\omega_{eq}$ and $\phi_k$. In Figure \ref{fig:branchpoint} we show the motion of the two poles near the collision point, thereby confirming that this is a branch point of the hydrodynamic dispersion relation.
Our results for $k_{eq}$ are consistent with those in \cite{Jansen:2020hfd}, which observed that $k_{eq}\rightarrow0$ as $T\rightarrow0$. As for $G_{\Pi\Pi}$, the collision that characterizes the breakdown of diffusive hydrodynamics in $G_{\varepsilon\varepsilon}$ changes qualitatively at sufficiently high $T$ \cite{Abbasi:2020ykq,Jansen:2020hfd}. 

Finally, we mention that it is manifest from Figure \ref{fig:RNallmodes} that even far outside the formal regime of applicability of hydrodynamics (i.e.~far past the first collision of the hydrodynamic pole), the system still supports modes whose dispersion relations are extremely well approximated by the quadratic approximation to diffusive hydrodynamics, with diffusivities given by \eqref{eq:RNdiffs}. We observed the same phenomenon in the neutral, translational symmetry-breaking model above, and it would be interesting to relate these observations to the results of \cite{Davison:2013bxa,Moitra:2020dal} on the hydrodynamic-like properties of zero temperature fluids.

\bigbreak

\section{SYK chain model}
\label{sec:SYKAppendix}

In this Appendix, we summarise the key features of the large-$q$ SYK chain model necessary for our purposes. We refer the reader to \cite{Choi:2020tdj} for more details. The Hamiltonian is
\begin{equation}
\begin{aligned}
H=&\,i^{q/2}\sum_{x=0}^{M-1}\left(\sum_{\substack{1\leq i_1<\ldots< i_q\leq N\\\\,\,\,}}J_{i_1\ldots i_q,x}\chi_{i_1,x}\ldots\chi_{i_q,x}\right.\\
&\,+\left.\sum_{\substack{1\leq i_1<\ldots<i_{q/2}\leq N \\ 1\leq j_1<\ldots<j_{q/2}\leq N}}J'_{i_1\ldots i_{q/2}j_1\ldots j_{q/2},x}\chi_{{i_1},x}\ldots\chi_{i_{q/2},x}\chi_{j_1,x+1}\ldots\chi_{j_{q/2},x+1}\right),
\end{aligned}
\end{equation}
where $\chi_{i,x}$ denote the $N$ Majorana fermions on site $x$ and $q$ is an even integer. The chain is composed of $M$ sites and the fermions obey periodic boundary conditions $\chi_{i,0}=\chi_{i,M}$. $J_{i_1\ldots i_q,x}$ and $J'_{i_1\ldots i_{q/2}j_1\ldots j_{q/2},x}$ are independent Gaussian random variables with zero mean and variances
\begin{equation}
\overline{J_{i_1\ldots i_q,x}^2}=\frac{(q-1)!}{N^{q-1}}J_0^2,\quad\quad\quad \overline{{J'_{i_1\ldots i_{q/2}j_1\ldots j_{q/2},x}}^2}=\frac{\left((q/2)!\right)^2}{qN^{q-1}}J_1^2.
\end{equation}

In the limit $N\gg q^2\gg 1$ with $\mathcal{J}_{0,1}^2=qJ_{0,1}^2/2^{q-1}$ fixed, an analytic expression for the energy density Green's function can be derived. From now on we fix the Euclidean time periodicity $1/T=2\pi$ and lattice spacing $a=1$. The Green's function is most conveniently expressed in terms of a dimensionless interaction strength $v$
\begin{equation}
\frac{\pi v}{\cos\left(\frac{\pi v}{2}\right)}=2\pi\sqrt{\mathcal{J}_0^2+\mathcal{J}_1^2},\quad\quad\quad\quad 0<v<1,
\end{equation}
and the relative strength of on-site and inter-site interactions $\gamma$
\begin{equation}
\gamma=\frac{\mathcal{J}_1^2}{\mathcal{J}_0^2+\mathcal{J}_1^2},\quad\quad\quad\quad 0<\gamma\leq 1.
\end{equation}
Our primary interest is in the limit of strong interactions $v\rightarrow1$, which restoring units of $T$ can be seen to correspond to $T\to0$.

The Fourier space energy density retarded Green's function is
\begin{equation}
\label{eq:SYKanalyticGreensfn}
G_{\varepsilon\varepsilon}(\omega,k)=-\frac{Nv}{2q^2}\left.\left(\partial_\theta\log\psi_n(\theta_v)+\tan\left(\frac{\pi v}{2}\right)\frac{h(h-1)}{2}\right)\right|_{n\rightarrow-i\omega+\epsilon},
\end{equation}
where
\begin{equation}
\begin{aligned}
\psi_n(\theta)=\,&\frac{\Gamma\left(1-\frac{h}{2}-\frac{n}{2v}\right)\sin\left(\frac{\pi h}{2}+\frac{\pi n}{2v}\right)\sin\left(\frac{n\pi}{2}\right)}{\Gamma\left(\frac{1}{2}-\frac{h}{2}+\frac{n}{2v}\right)}\cos(\theta)\sin(\theta)^h \,_2F_1\left(\frac{1+h-\frac{n}{v}}{2},\frac{1+h+\frac{n}{v}}{2},\frac{3}{2},\cos^2\theta\right)\\
\,&+\frac{\Gamma\left(\frac{1}{2}-\frac{h}{2}-\frac{n}{2v}\right)\cos\left(\frac{\pi h}{2}+\frac{\pi n}{2v}\right)\cos\left(\frac{n\pi}{2}\right)}{2\Gamma\left(1-\frac{h}{2}+\frac{n}{2v}\right)}\sin(\theta)^h \,_2F_1\left(\frac{h-\frac{n}{v}}{2},\frac{h+\frac{n}{v}}{2},\frac{1}{2},\cos^2\theta\right),
\end{aligned}
\end{equation}
and
\begin{equation}
h=\frac{1}{2}\left(1+\sqrt{9+4\gamma\left(\cos(k)-1\right)}\right),\quad\quad\quad\quad \theta_v=\frac{\pi}{2}\left(1-v\right).
\end{equation}
This Green's function exhibits a hydrodynamic diffusion mode for all $v$, with diffusivity
\begin{equation}
D_\varepsilon=\frac{\gamma v}{12}\left(\pi v\tan\left(\frac{\pi v}{2}\right)+2\right),
\end{equation}
in these units.

The emergence of a series of long-lived infra-red modes in the limit of strong interactions may be seen by taking the $v\rightarrow1$ limit of \eqref{eq:SYKanalyticGreensfn} at fixed $\omega,k$. In this limit
\begin{equation}
G_{\varepsilon\varepsilon}\propto(1-v)^{2h-2}\frac{\Gamma\left(\frac{1}{2}-h\right)\Gamma\left(h-i\omega\right)}{\Gamma\left(\frac{1}{2}+h\right)\Gamma\left(1-h-i\omega\right)},
\end{equation}
up to a $\omega$-independent proportionality constant, and neglecting contact terms. This expression is just the infra-red Green's function $\mathcal{G}_{IR}$ defined in equation \eqref{eq:IRgreensfn} (in the units $2\pi T=1$), with infra-red scaling dimension
\begin{equation}
\Delta(k)=h=\frac{1}{2}\left(1+\sqrt{9+4\gamma\left(\cos(k)-1\right)}\right).
\end{equation}
Therefore $\Delta(0)=2$ sets the equilibration time of this system.

To obtain the local equilibration data for Figure \ref{fig:SYKplotmaintext} in the main text, we numerically evaluated the locations of poles of the expression \eqref{eq:SYKanalyticGreensfn} for $G_{\varepsilon\varepsilon}$, and from these identified the location of the collision that signals the breakdown of diffusive hydrodynamics. Note that $k$ and $\gamma$ enter the Green's function only through the combination $h$ and thus the poles can be parameterised by $\omega(h)$. As a consequence of this, the collision frequency $\omega_{eq}$ is independent of $\gamma$.

\vskip1cm

%\bibliography{univIRrelax-biblio}
\end{document}